\PassOptionsToPackage{table,xcdraw}{xcolor}
\documentclass[sigplan,screen]{acmart}

\usepackage{tikz}
\usepackage{amsmath}

\usepackage[utf8]{inputenc}
\usepackage{url}
\usepackage[T1]{fontenc}
\usepackage{graphicx}
\usepackage{listings}
\usepackage{epstopdf}
\usepackage{balance}
\usepackage{color}
\usepackage{subfigure}
\usepackage{xspace}
\usepackage{latexsym}
\usepackage{wrapfig}
\usepackage{setspace}
\usepackage{blindtext}
\usepackage{amsfonts}
\usepackage{multirow}
\usepackage{pifont}
\usepackage{mathrsfs}
\usepackage{mathtools}
\usepackage{relsize}
\usepackage{lipsum}

\usepackage{graphics}

\usepackage{ulem}

\usepackage{comment}
\usepackage{footnote}
\usepackage{pifont}
\usepackage{makecell}
\usepackage{multirow}
\usepackage{booktabs}
\usepackage{hhline}
\usepackage{adjustbox}

\usepackage{cleveref}
\crefformat{section}{\S#2#1#3}

\usepackage[most]{tcolorbox}
\definecolor{light-gray}{gray}{0.95}
\newtcolorbox{observation}{
    center,
    width=\linewidth,
    colframe=light-gray,
    colback=light-gray
}

\newcommand*\circled[1]{\tikz[baseline=(char.base)]{
            \node[shape=circle,fill,inner sep=1pt] (char) {\textcolor{white}{#1}};}}

\newcommand{\proj}{\textsf{NotebookOS}} 
\newcommand{\reservation}{\textsf{Reservation}} 
\newcommand{\FCFS}{\textsf{Batch}} 
\newcommand{\projlcp}{\textsf{NotebookOS (LCP)}}
\newcommand{\phillytrace}{\textsf{PhillyTrace}}

\newcommand{\platformtrace}{\textsf{AdobeTrace}} 
\newcommand{\idltplatform}{\textsf{Adobe research cluster}}

\newcommand{\alibabatrace}{\textsf{AlibabaTrace}}

\newif{\ifSubmit}
\newif{\ifFinal}
\newif{\ifDraft}

\Finaltrue 
\Submittrue 

\ifSubmit
\newcommand{\yuec}[1]{}
\newcommand{\todo}[1]{}
\newcommand{\bencomment}[1]{}
\newcommand{\jycomment}[1]{}
\newcommand{\maicomment}[1]{}
\newcommand{\rzcomment}[1]{}
\newcommand{\added}[1]{#1}
\else
\newcommand{\added}[1]{\noindent\textcolor{brown}{#1}}
\newcommand{\todo}[1]{\noindent\textcolor{blue}{#1}}
\newcommand{\yuec}[1]{\noindent\textcolor{red}{Yue: #1}}
\newcommand{\jycomment}[1]{\noindent\textcolor{orange}{\bf Jingyuan: #1}}
\newcommand{\bencomment}[1]{\noindent\textcolor{blue}{\bf Ben: #1}}
\newcommand{\maicomment}[1]{\textcolor{brown}{\textbf{Mai: #1}}}
\newcommand{\rzcomment}[1]{\textcolor{teal}{\textbf{Lukas: #1}}}
\fi

\ifDraft
\newcommand{\todocomment}[1]{\textcolor{purple}{#1}}
\newcommand{\addcomment}[1]{\textcolor{red}{#1}}
\newcommand{\diffcomment}[2]{\textcolor{orange}{#1}}
\newcommand{\delcomment}[1]{\textcolor{orange}{\sout{#1}}}
\else
\newcommand{\todocomment}[1]{}
\newcommand{\addcomment}[1]{#1}
\newcommand{\diffcomment}[2]{#1}
\newcommand{\delcomment}[1]{}
\fi

\definecolor{linkblue}{HTML}{005697}

\lstdefinestyle{bash_cmd_style}{
    language=bash,
    basicstyle=\scriptsize\ttfamily,
    breaklines=true,
    columns=fixed
}

\copyrightyear{2026}
\acmYear{2026}
\setcopyright{cc}
\setcctype{by}
\acmConference[ASPLOS '26]{Proceedings of the 31st ACM International
Conference on Architectural Support for Programming Languages and Operating
Systems, Volume 1}{March 22--26, 2026}{Pittsburgh, PA, USA}
\acmBooktitle{Proceedings of the 31st ACM International Conference on
Architectural Support for Programming Languages and Operating Systems,
Volume 1 (ASPLOS '26), March 22--26, 2026, Pittsburgh, PA, USA}
\acmDOI{10.1145/3760250.3762230}
\acmISBN{979-8-4007-2165-6/2026/03}

\begin{document}

\pagenumbering{gobble}

\date{}

\title{{\proj}: A Replicated Notebook Platform for Interactive Training with On-Demand GPUs} 

\author{Benjamin Carver}
\orcid{0000-0002-1574-9300}
\affiliation{%
  \institution{{\it George Mason University}}
  \city{Fairfax} 
  \state{Virginia}
  \country{USA} 
}
\email{bcarver2@gmu.edu}

\author{Jingyuan Zhang}
\orcid{0000-0001-9581-1807}
\affiliation{%
  \institution{{\it George Mason University}}
  \city{Fairfax} 
  \state{Virginia}
  \country{USA} 
}
\email{jzhang33@gmu.edu}

\author{Haoliang Wang}
\orcid{0000-0003-2963-8721}
\affiliation{%
  \institution{{\it Adobe Research}}
  \city{San Jose} 
  \state{California}
  \country{USA} 
}
\email{hawang@adobe.com}

\author{Kanak Mahadik}
\orcid{0009-0007-5176-662X}
\affiliation{%
  \institution{{\it Adobe Inc}}
  \city{San Jose} 
  \state{California}
  \country{USA} 
}
\email{mahadik@adobe.com}

\author{Yue Cheng}
\orcid{0000-0003-1695-4864}
\affiliation{%
  \institution{{\it University of Virginia}}
  \city{Charlottesville} 
  \state{Virginia}
  \country{USA} 
}
\email{mrz7dp@virginia.edu}
\authornote{Corresponding author}

\begin{abstract}

Interactive notebook programming is universal in modern ML and AI workflows, with interactive deep learning training (IDLT) emerging as a dominant use case. To ensure responsiveness, platforms like Jupyter and Colab reserve GPUs for long-running notebook sessions, despite their intermittent and sporadic GPU usage, leading to extremely low GPU utilization and prohibitively high costs. In this paper, we introduce {\proj}, a GPU-efficient notebook platform tailored for the unique requirements of IDLT. {\proj} employs replicated notebook kernels with Raft-synchronized replicas distributed across GPU servers. To optimize GPU utilization, {\proj} oversubscribes server resources, leveraging high inter-arrival times in IDLT workloads, and allocates GPUs only during active cell execution. It also supports replica migration and automatic cluster scaling under high load. Altogether, this design enables interactive training with minimal delay. In evaluation on production workloads, {\proj} saved over 1,187 GPU hours in 17.5 hours of real-world IDLT, while significantly improving interactivity.

\end{abstract}



\maketitle

\section{Introduction}
\label{sec:introduction}

Interactive notebook programming is universal in modern ML (machine learning) and AI (artificial intelligence) workflows. 
Software such as Jupyter Notebook~\cite{jupyter_notebook} and Google Colab~\cite{colaboratory} provides a user-friendly, interactive, web-based programming interface, 
and therefore, they have become the \emph{de facto} interface for interactive data-driven programming, e.g., data analytics, data science, and AI/ML, spanning almost every science and engineering domain. 
Take Jupyter Notebook as an example. 
To date, Jupyter Notebook has been used by millions of users world wide~\cite{future_jupytercon23} in education~\cite{jupyter_edu_book, jupyter_edu_projs}, scientific research~\cite{nasa_jupytercon23, databook_py_demo}, and collaborations~\cite{jupyter_rt_collaboration, collaborate_jupytercon23}. 
A series of industry initiatives~\cite{netflix_notebooks, paypal_notebooks, bloomberg_notebooks}, workshops~\cite{jupyter_community_workshop19, jupytercon}, and a vast array of resources, projects, and libraries~\cite{awesome_jupyter} underscore the growing demand for using Notebooks in data-driven jobs and tasks. 
In addition, all major cloud providers, along with an increasing number of startups, now offer commercial Notebook services~\cite{jupyter_aws, sagemaker_notebooks, sagemaker_studio_lab,  gcp_vertex_notebooks, microsoft_notebooks, naas, lentiq, adobe_experience_platform}.

Traditionally, GPU-based, batch deep learning training (BDLT) features batch-style, long-running workloads~\cite{tiresias_nsdi19, gandiva_osdi18, optimus_eurosys18, themis_nsdi20, philly_trace_atc19, alibaba_cluster_2020_trace_nsdi22} that require uninterrupted access to GPU resources over long periods of time, such as large-scale model training. These traditional systems focus on maximizing throughput and job-completion time (JCT), often prioritizing long-running tasks that can afford latency and resource contention. 
In contrast, notebook-oriented, \emph{interactive deep learning training (IDLT)} workloads---which consist of tasks like model and program debugging,  dynamic/iterative model adjustments, hyperparameter tuning---demand a different set of performance characteristics. This set includes low latency, responsiveness, high interactivity, and efficient management of many short-running tasks. The short-lived tasks that dominate IDLT workloads are executed within the context of long-lived user sessions, such as Jupyter Notebook sessions.

To guarantee interactivity, today's notebook services typically provision and reserve GPU resources during the entire lifetime of a Notebook session. Although these sessions are long-running, cell executions are fragmented: IDLT workloads exhibit intermittent, sporadic, and often transient GPU usage. 
Users simply spend more time developing a notebook than they do executing the notebook's tasks. As a result, notebook sessions spend a majority of their time not using the reserved GPUs, leading to \emph{extremely low GPU utilization and prohibitively high cost}.

We build {\proj}, a  \emph{first-of-its-kind} \textit{notebook platform} designed for the unique requirements of IDLT. {\proj} is distinct in both purpose and design from traditional \textit{GPU cluster computing} systems~\cite{gandiva_osdi18, themis_nsdi20, tiresias_nsdi19, gavel_osdi20, antman_osdi20, pollux_osdi21, sia_sosp23}. 
{\proj} prioritizes interactivity, low response time, and efficient management of numerous short-lived GPU tasks. By addressing these unique needs of users actively working in notebooks, {\proj} realizes a novel notebook platform that uses fine-grained, flexible GPU allocation to effectively meet the demands of short, dynamic GPU tasks.

At {\proj}'s core is a collection of novel techniques and design tenets that enable the efficient support of IDLT workloads. {\proj} uses a pluggable notebook task scheduling and placement mechanism,  
whose default policy is designed to maximize interactivity. 
\addcomment{Specifically, {\proj} uses a replicated kernel design, where each notebook kernel is replicated across multiple GPU servers. 
Any replica can execute CPU or GPU tasks. Small kernel state is synchronized across replicas using Raft~\cite{raft_atc14}, while large objects are asynchronously replicated via a distributed data store. This design decouples GPU allocation from notebook sessions: GPUs are dynamically allocated and assigned to one of the replicas only during code execution, maximizing the chances of immediate training upon code submission. 
To further improve GPU utilization, {\proj} oversubscribes GPU resources on each server, leveraging the high task inter-arrival time (IAT) observed in IDLT workloads.} This maximizes kernel availability and minimizes wait times during user code submission.

In summary, this paper makes the following contributions:

\begin{enumerate}
\vspace{-2pt}
    \item Characterization of an important and under-studied class of DL workloads called IDLT. 
    \item Design and implementation of {\proj} tailored for the unique requirements of notebook IDLT. 
    \item Comprehensive evaluation of {\proj} using prototype and simulation on production IDLT workloads. 
\end{enumerate}
\vspace{-2pt}

{\proj} aims 
to pave the way for more efficient AI compute via on-demand GPUs.  
{\proj} is available at: \href{https://github.com/ds2-lab/NotebookOS}{\textcolor{blue}{https://github.com/ds2-lab/NotebookOS}}.

\vspace{-2pt}
\section{Background and Motivation}
\label{sec:background}

This section presents an overview of relevant background and a workload analysis that motivates the need for a new platform tailored for notebook IDLT workloads.

\begin{figure}[t]
\centering
\includegraphics[width=0.425\textwidth]{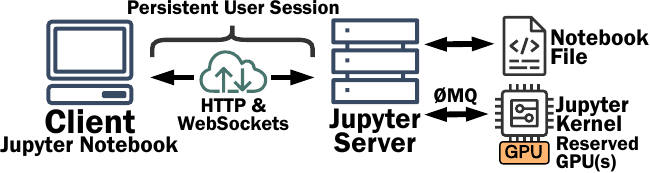}
\vspace{-8pt}
\Description[Client-server Jupyter Notebook architecture.]{Client-server Jupyter Notebook architecture.}
\caption{Client-server Jupyter Notebook architecture.} 
\label{fig:generic_jupyter_notebook_arch}
\vspace{-14pt} 
\end{figure}

\vspace{-2pt} 
\subsection{Jupyter Notebook}

Jupyter Notebook~\cite{jupyter_notebook} is an open-source, web-based interactive development environment (IDE) designed for interactive programming.
Jupyter Notebook is most commonly used with Python and can be used with other languages~\cite{jupyter_available_kernels}, such as R~\cite{jupyter_irkernel} and Julia~\cite{jupyter_ijulia}. Jupyter Notebook has become the \textit{de facto} development environment for data-driven programming: ML/AI, data science, and data analytics. This is due in large part to its versatility, ease-of-use, and interactivity.  

\noindent\textbf{Terminology.} \emph{Notebooks} are interactive documents that combine code, text for explaining the code, and visualizations for displaying the code's inputs and outputs. 
These elements are contained within the \emph{cells} of the notebook, which are the basic ``units'' of a notebook. When a user runs code in some cell(s), the code is sent to a separate process, called a \emph{kernel}, which resides either on the user's local computer or on a remote server. 
The kernel executes the code and returns results back to the notebook document where they are displayed in the cell(s). 
We refer to this process as a \emph{cell task execution}. A \emph{notebook session} is a persistent, working instance of a Jupyter Notebook environment where the state of variables, imports, and other execution context are maintained and reused by the associated kernel. 
Finally, we define an \emph{IDLT task} as a cell task execution of GPU operations---such as model training and  inference---during an \emph{interactive} workload involving debugging and testing, initial model design, 
and other similar activities. See Figure~\ref{fig:generic_jupyter_notebook_arch} for an illustration.

\noindent\textbf{Notebook-as-a-Service}
In recent years, Jupyter Notebook-as-a-Service (NaaS) platforms have emerged, providing 
fully-managed, configurable, and scalable notebook environments. 
NaaS enables users to focus on coding, analysis, and research by abstracting infrastructure complexities such as server management and resource scaling. Popular Jupyter-based NaaS platforms include Google Colab~\cite{colaboratory}, Amazon SageMaker Studio~\cite{aws_sagemaker_studio}, and CoCalc~\cite{cocalc}. 

\subsection{Interactive Deep Learning Training}

IDLT is a popular type of applications performed using Jupyter Notebook. Examples of IDLT include interactive model debugging, exploratory data analysis, hyperparameter tuning, and fast iteration of ideas, among others. Notebook-based IDLT is typically deployed and executed on large-scale GPU clusters~\cite{colaboratory,naas, azure_ml, aws_sagemaker_studio} and demands high interactivity/low response time for optimal performance.

\begin{figure*}[t]
\begin{center}
\subfigure[Task duration CDF.] {
\includegraphics[height=0.165\textwidth]{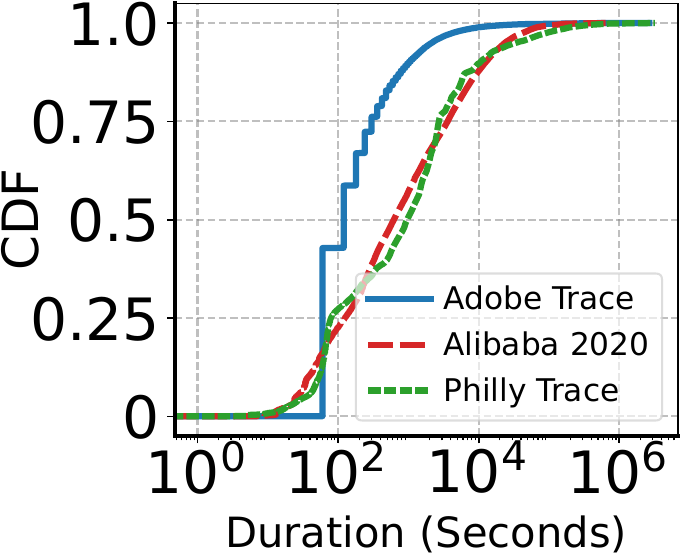}
\label{fig:training_task_duration_cdf}
}
\hspace{-6pt}
\subfigure[Inter-arrival time CDF.] {
\includegraphics[height=0.165\textwidth]{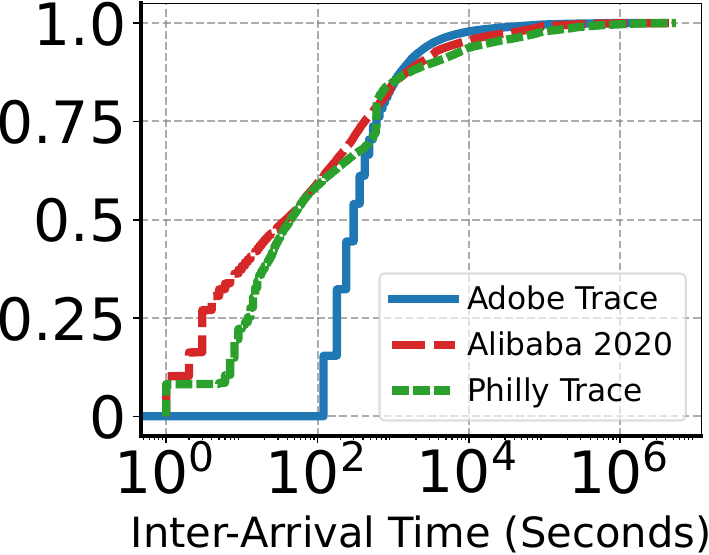}
\label{fig:training_task_iat_cdf}
}
\hspace{-6pt}
\subfigure[GPU util. CDF ({\platformtrace}).] {
\includegraphics[height=0.165\textwidth]{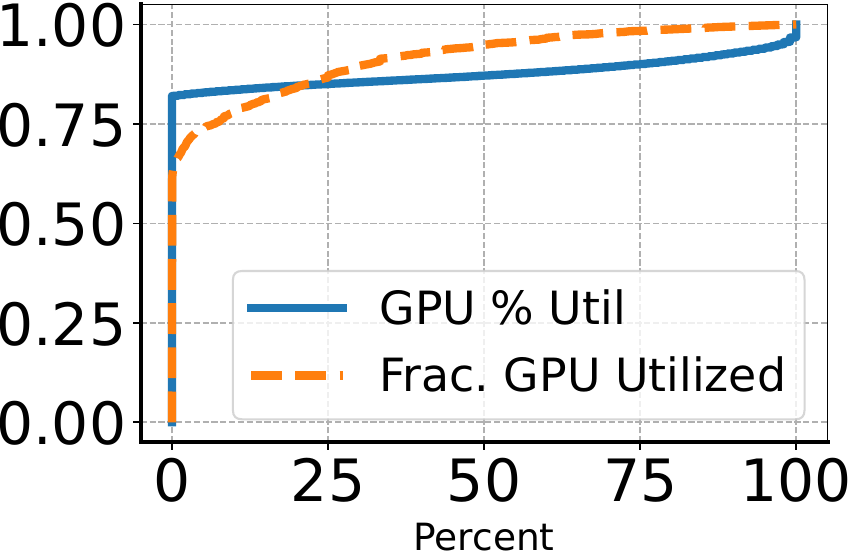}
\label{fig:gpu_util_cdfs}
}
\hspace{-6pt}
\subfigure[GPU \& CPU usage ({\platformtrace}).] {
\includegraphics[height=0.17\textwidth]{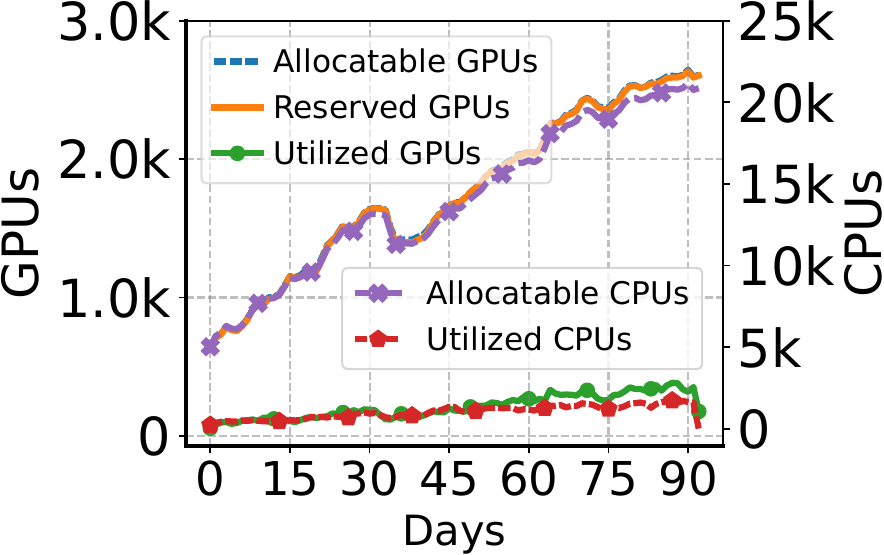}
\label{fig:gpu_usage_timeline}
}
\vspace{-8pt}
\Description[Workload characteristics of three representative GPU cluster traces.]{Workload characteristics of three representative GPU cluster traces.}
\caption{Workload characteristics of three representative GPU cluster traces.}
\label{fig:workload_analysis_combined}
\vspace{-8pt}
\end{center}
\end{figure*}

IDLT workloads exhibit intermittent and highly variable GPU usage patterns. In order to guarantee high interactivity, modern NaaS platforms typically provision and reserve GPU resources for the entire lifetime of the notebook session (Figure~\ref{fig:generic_jupyter_notebook_arch}). 
As a result, notebook sessions tend to spend a majority of their lifetime \textit{not} using their reserved GPU resources: while users are debugging their code or performing other similar tasks, the GPUs are idle, and thus resources are wasted. This pattern ultimately results in extremely low GPU utilization and prohibitively high monetary costs.

\vspace{-2pt}
\subsection{Real-World IDLT Workload Analysis}
\label{subsec:workload_analysis}

Real-world IDLT workloads exhibit distinct behaviors compared to traditional long-running training jobs. 
In order to better understand these workloads and their requirements, we present a workload analysis of the production notebook IDLT traces collected from an Adobe internal research cluster~\cite{adobe}.
{\idltplatform} hosts containerized notebook sessions within AWS EC2 GPU virtual machine (VM) instances. {\idltplatform} also manages the GPU resource allocations for the notebook sessions. 
To deploy a notebook container, users specify how many GPUs should be allocated. 
Once launched, GPUs are bound to the notebook container, and the user can interact with the back-end notebook kernel (i.e., an IPython process) via a web-based Jupyter Notebook interface. 

The IDLT traces ({\platformtrace}) include 545,467 individual training events. 
We focus on a representative subset spanning June 1--August 31. The sample granularity of this trace is 15 seconds. %

For context, we compare {\platformtrace} with two other popular, publicly available GPU cluster traces. First, we use the Alibaba GPU Cluster'20 trace~\cite{alibaba_cluster_2020_trace_nsdi22} ({\alibabatrace}), which captures both training and inference jobs running state-of-the-art ML algorithms from July to August 2020. Data were collected from a large production cluster with 6,500 GPUs across 1,800 servers. We also include the Philly GPU trace~\cite{philly_trace_atc19} ({\phillytrace}), a 6.6GB representative subset of first-party deep learning training (DLT) workloads from Microsoft's internal Philly clusters. {\phillytrace} consists of 117,325 jobs collected from a cluster of 2,490 GPUs distributed across 552 servers between August 7 -- December 22, 2017. 

We chose these two traces for comparison because they represent typical DLT workloads, which are characterized by long-running training tasks that are scheduled by a batch scheduler and span several hours. As we show in our analysis, the workload captured by {\platformtrace} differs considerably from that of {\phillytrace} and {\alibabatrace}.

\vspace{-2pt}
\subsubsection{Training Task Duration}
\label{subsubsec:_workload_analysis_training_task_duration}

Figure~\ref{fig:training_task_duration_cdf} displays a cumulative distribution function (CDF) of the training task durations, in seconds, of DLT tasks submitted by users to the {\idltplatform}. The data used to plot the CDF were taken from the June, July, and August 2021 trace dataset. 50\% of user-submitted tasks are 2 minutes or less, while 75\% are 5 minutes or less. 90\% of training tasks are 17 minutes or less, 95\% are 36 minutes or less, and the duration of the 99$^{th}$ percentile training task is 182 minutes (i.e., 3.0$\bar{3}$ hours). Based on this, it is clear that an overwhelming majority of training tasks are short-lived. %

The {\platformtrace} workload consists of a significantly high fraction of \textit{short} training tasks compared to both {\phillytrace} and {\alibabatrace}. %
Specifically, the 50$^{th}$ percentile of task durations is 120 seconds (2 minutes), 621 seconds (10.35 minutes), and 957 seconds (15.95 minutes) for {\platformtrace}, {\phillytrace}, and {\alibabatrace}, respectively.
This observation highlights the difference between traditional BDLT workloads (i.e., {\phillytrace} and {\alibabatrace}) and notebook IDLT workloads (i.e., {\platformtrace}). Note that a meaningful comparison of task durations less than 15 seconds is impossible due to the 15-second granularity of {\platformtrace}. 

\noindent\textbf{\textit{Observation 1:}}
 \emph{IDLT workloads contain a large percentage of very short tasks---with 75\% of tasks completing in 5 minutes or less---significantly shorter than traditional BDLT workloads.}

\vspace{-2pt}
\subsubsection{Training Task Inter-Arrival Time (IAT)}
\label{subsubsec:_workload_analysis_training_task_iat}

Figure \ref{fig:training_task_iat_cdf} shows a CDF plot of the IATs of training tasks from the {\idltplatform} trace. For all three traces, the IATs were measured within each user session \textit{independently}, rather than the cluster-wide IATs of tasks submitted by any active user session, to ensure a fair comparison. From this graph, we can see that the 50$^{th}$ percentile of task IATs is 300 seconds (5 minutes), 44 seconds, and 38 seconds for {\platformtrace}, {\phillytrace}, and {\alibabatrace}, respectively.

The observed {\platformtrace} IATs align with the nature of IDLT workloads, where users \emph{intermittently} make small, debugging-like changes to their models or Python code before submitting a \emph{short} cell task to test the changes. In particular, {\platformtrace} exhibits significantly longer IATs for more than 50\% of tasks compared to {\phillytrace} and {\alibabatrace}. This is likely due to {\platformtrace} users engaging in more extensive interactive cycles, iteratively testing and refining their code before each submission, rather than executing fully tested, production-ready DLT tasks scheduled by a batch GPU scheduler.  

\noindent\textbf{\textit{Observation 2:}}
\emph{IDLT tasks are submitted less frequently, with 75\% having an IAT of at most 480 seconds (8 minutes), as users do not submit concurrent tasks and often make iterative modifications and tests after a task completes.}

\subsubsection{Cluster GPU Utilization}
\label{subsubsec:_workload_analysis_gpu_utilization}

Figure~\ref{fig:gpu_util_cdfs} plots two different but related data series whose units are a percentage. The first series, shown as the solid blue line, is a CDF of cluster GPU utilization, measured every 15 seconds across all reserved GPUs throughout the trace's duration. The second, shown as a dashed orange line, is a CDF of the percentage of each session's lifetime during which allocated GPUs were actively utilized: a 50\% value means GPUs were actively utilized for half of the time that their sessions were active. 

One key observation from Figure~\ref{fig:gpu_util_cdfs} is that the reserved GPU resources were idle over 81\% of the time. Moreover, 
nearly 70\% of GPUs were \textit{completely idle} throughout the entire lifetime of the notebook session to which they were assigned. (The lifetime of the notebook session begins when the associated container is provisioned and ends when that container is terminated.) Between 74\% and 75\% of user sessions actively used their allocated GPUs at most 5\% of the time, and 90\% of sessions only used their allocated GPU resources at most 31.13\% of the time.

Figure~\ref{fig:gpu_usage_timeline} plots the number of GPUs (and CPUs) that are actively utilized compared to the number of reserved GPUs (CPUs). There is a significant gap between the number of utilized GPUs (CPUs) and the number of reserved GPUs (CPUs). By the conclusion of the 3-month period, only about $15\%$ of all reserved GPUs are actively utilized. 

These results highlight the urgent demand for a fundamentally new approach to notebook IDLT resource management to address the high degree of wasted resources.  

\noindent\textbf{\textit{Observation 3:}} 
\emph{Notebook users often underutilize their allocated GPU resources during IDLT: reserved GPU resources were idle over 81\% of the time.}

\vspace{-6pt} 
\subsection{Insight and Challenges}

\noindent\textbf{Key Insight.}
IDLT workloads fundamentally differ from traditional BDLT workloads. This difference arises in task duration, submission frequency (i.e., IATs), and the need for user feedback or user input. During conventional BDLT workloads, ML models are trained over long periods of time---many hours or even days--with minimal intervention, except in case of failures~\cite{alibaba_cluster_2020_trace_nsdi22, gemini_sosp23}. In contrast, a key insight is that IDLT workloads are characterized by intermittent and low GPU usage patterns. This is because, a user's Jupyter Notebook session has GPUs bound to it from the beginning, even though the user may spend a lot of time doing work (coding, debugging, etc) that does not require a GPU and only occasionally executing training tasks that require a GPU.

\noindent\textbf{Challenges.}
This workload analysis highlights key challenges that must be addressed to design an efficient system for supporting notebook-based IDLT workloads.

\noindent$\bullet$ \textbf{C1: Resource Utilization.}
Modern notebook platforms typically reserve GPUs, delegating their management to users~\cite{jupyterhub, colaboratory}. Users must manually request GPUs, CPUs, and host memory, as well as start and shut down their notebook sessions. To avoid the hassle of termination, many leave notebooks idling until the provider reclaims resources~\cite{jupyter_culling_idle_kernel}. 

This behavior leads to extremely low GPU utilization, as discussed in \cref{subsec:workload_analysis}.

\noindent$\bullet$ \textbf{C2: Interactivity.} 
Jupyter Notebook offers a user-friendly, interactive interface for development, execution, and debugging within a web browser. 
Some NaaS providers, like Azure Machine Learning~\cite{azure_ml}, link the front-end notebook interface to a backend batch scheduler. However, this setup introduces high (container startup and batch queueing) delays, degrading interactivity and user experience~\cite{ncar_papermill, osc_jupyter_batchconnect}. 

\noindent$\bullet$ \textbf{C3: Resource Elasticity.}
Low resource utilization stems from static GPU over-provisioning and a lack of elastic resource management in NaaS platforms. {\idltplatform}, operating roughly 12k V100 GPUs across 3k {\texttt\small{p3.16xlarge}} EC2 instances, incurs roughly \$18.3 million in monthly costs even with long-term reservation discounts. 
This high TCO (total cost of ownership), combined with the extremely low resource utilization, highlights the urgent need for a GPU-efficient notebook platform that can flexibly adapt based on real-time workload requirements.

\vspace{-6pt}
\section{{\proj} Design}
\label{sec:design}

\noindent\textbf{Workload Requirements.} 
We present {\proj}, a GPU-efficient notebook platform to address the challenges outlined in \cref{sec:background}. 
Traditional GPU cluster schedulers are designed with long-running GPU-intensive tasks in mind. 
They often prioritize
job completion time (JCT)~\cite{gandiva_osdi18, optimus_eurosys18, tiresias_nsdi19}, throughput~\cite{gavel_osdi20}, and fairness~\cite{themis_nsdi20, gavel_osdi20} over responsiveness and interactivity in that they allocate GPUs in a way that balances the competing demands of long-running jobs. Therefore, they 
are poorly-suited for notebook IDLT. Whereas slight delays may have been acceptable before, even short delays can severely impact the user experience when performing highly interactive 
tasks.
\emph{In {\proj}, the primary goal is to provide users with fast GPU access for \textit{interactive} sessions. This demands flexible and low-latency resource allocation.}

Another requirement is \emph{user sessions}, which traditional GPU schedulers overlook. They treat each task as an independent request competing for resources. 
This is a poor match for notebook IDLT where users  typically work within a long-running, \emph{stateful} notebook session, incrementally modifying the Python code, and intermittently submitting  GPU tasks. 
{\proj} addresses this gap by introducing mechanisms to efficiently schedule short GPU tasks within an ongoing session, enabling smarter resource allocation based on real-time user needs rather than isolated batch jobs. 

\noindent\textbf{Design Choices.}
{\proj} introduces several novel techniques and design decisions that harmonize between long-running, stateful notebook sessions and elastic GPU allocations for short-lived notebook tasks%
, including:

\noindent$\bullet$ \textbf{Distributed Notebook Kernels:}
{\proj} adopts a novel distributed notebook design, where each logical Jupyter kernel is mapped to a {\proj} \textit{distributed} kernel, each of which consists of $N$ kernel replicas scheduled across {\proj}'s GPU server cluster. 
This distributed kernel is designed to oversubscribe the under-utilized GPU resources. Its benefit is twofold. 
(1) Replicated kernels increase the likelihood that at least one replica will have immediate access to GPU resources upon code submission ({\bf C2}). 
(2)~Significantly improved resource utilization via GPU oversubscription ({\bf C1}).

\noindent$\bullet$ \textbf{State Machine Replication:}
{\proj}'s kernel replicas use the Raft consensus protocol~\cite{raft_atc14} for leader elections and 
State Machine Replication (SMR). SMR enables efficient, transparent CPU state synchronization between the kernel replicas that occurs \textit{off the critical path}. This design is motivated by the observation that interactive Python applications frequently use global variables to store intermediate state. These variables can be synchronized using SMR to ensure state changes are seen by all kernel replicas ({\bf C2}). 

\noindent$\bullet$ \textbf{Dynamic GPU Binding:} 
{\proj} does not exclusively commit GPU resources to notebook kernel replica containers long-term. Instead, GPUs are exclusively allocated to kernel replica containers only while user-submitted cell execution tasks are actively-running. Once a task completes, the GPUs are released, enabling them to be allocated to another co-located kernel replica. 
This approach allows {\proj} to adapt to fluctuating resource demands, allocating GPUs based on a session’s \emph{current} needs rather than its peak requirements. Users performing smaller-scale, less demanding tasks can request fewer GPUs, improving fine-grained resource efficiency ({\bf C1 and C3}). 

\noindent$\bullet$ \textbf{Transparent GPU State Checkpointing:}
Instead of performing SMR, {\proj} %
checkpoints large GPU state objects (model parameters and training datasets) asynchronously and separately to a distributed storage system, anticipating that the destination replica may become active ({\bf C2}).

\vspace{-10pt}
\subsection{{\proj} Overview}
\label{subsec:design_overview}

\begin{figure}[t]
\centering
\includegraphics[width=0.485\textwidth]{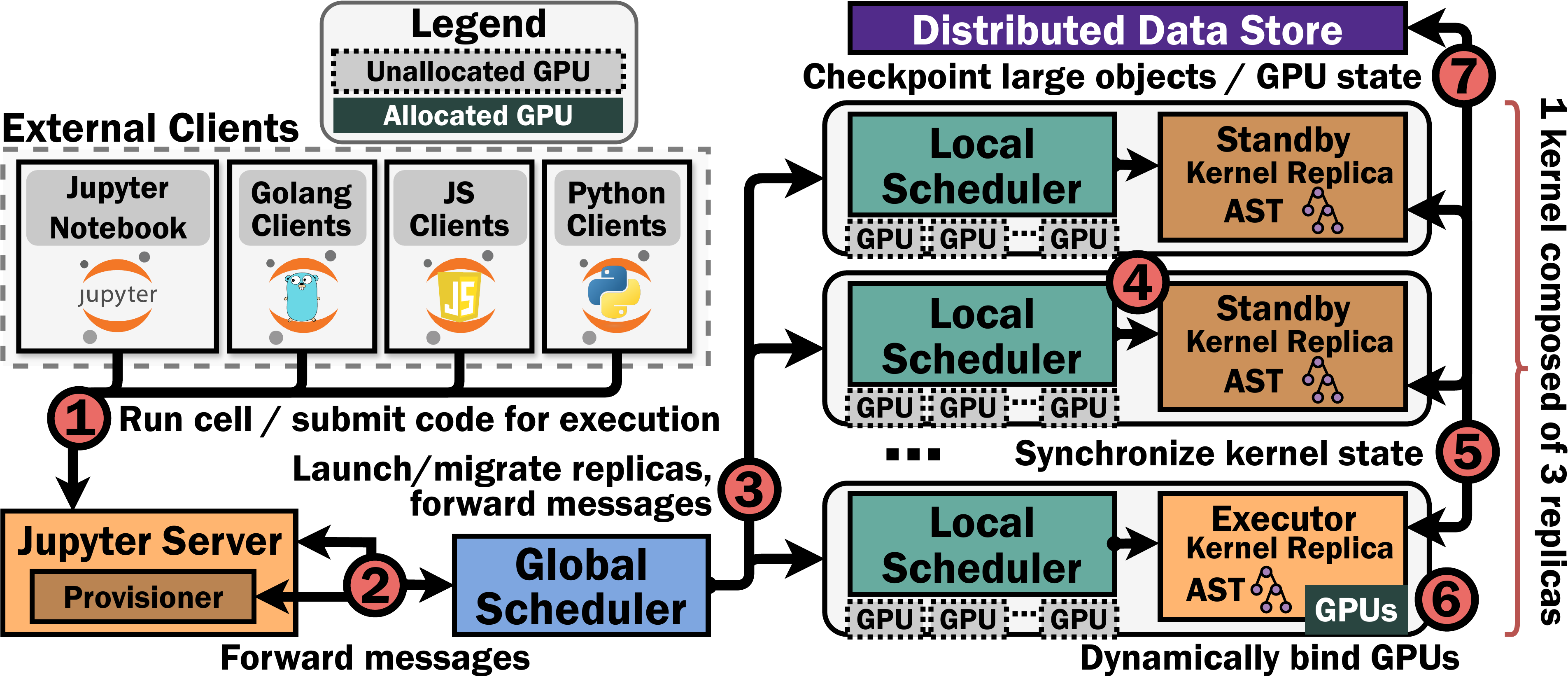}
\vspace{-14pt}
\caption{Architecture overview of {\proj}.} 
\Description[Architecture of {\proj}.]{Architecture of {\proj}.}
\label{fig:prototype_architecture}
\vspace{-16pt}
\end{figure}

Figure~\ref{fig:prototype_architecture} depicts the architecture of {\proj}. {\proj} consists of five core components, described below. 

\noindent\textbf{{\proj} Clients.}
Like traditional Jupyter environments, {\proj} clients interface with {\proj} by sending notebook operations via HTTP or WebSocket messages to {\proj}'s Jupyter Server (Step~\circled{1} of Figure~\ref{fig:prototype_architecture}).

\noindent\textbf{Jupyter Server.}
The Jupyter Server provides the core services, serving APIs and REST endpoints used by Jupyter web apps like Jupyter Notebook~\cite{jupyter_notebook} and JupyterLab~\cite{jupyter_lab_collaborate}. 

\noindent\textbf{Global Scheduler.}
The Global Scheduler is responsible for creating distributed kernels and initiating the provisioning of their kernel replicas (an operation that is delegated to another component of {\proj}, the Local Scheduler). It performs a majority of the book-keeping required by {\proj}, including managing the allocation of compute resources (i.e., CPUs, host memory, GPUs) to Distributed Kernel replicas, handling failures, kernel replica migration, and auto-scaling. 

Additionally, the Global Scheduler is responsible for routing messages from Jupyter clients to the appropriate Distributed Kernel replicas. Each Jupyter message, sent from a client and forwarded by the Jupyter Server (Step~\circled{2}), contains information such as the unique identifier of the target kernel. The Global Scheduler inspects this information and then routes the message to the replicas of the target kernel. Note that messages are first forwarded to the target kernel replica's Local Scheduler (Step~\circled{3}).

\noindent\textbf{Local Scheduler.}
{\proj} deploys a Local Scheduler on each GPU server. It forwards messages from the Global Scheduler to the target kernel replica running on the Local Scheduler's server. It is also responsible for provisioning and managing the containers in which the kernel replicas run. It also ensures proper cleanup upon kernel termination. 
Upon receiving a message from the Global Scheduler, the Local Scheduler routes it to the target kernel replica (Step~\circled{4}).  

\noindent\textbf{Distributed Kernel.}
{\proj}'s Distributed Kernel consists of three replicas, which use the Raft SMR protocol to replicate the CPU-memory state of the IPython process (Step~\circled{5} and \cref{subsubsec:state_synchronization}).
{\proj} designs a lightweight executor election protocol to elect a proper replica with sufficient GPU resources for running GPU tasks (Step~\circled{6} and \cref{subsubsec:kernel_raft_protocol}). 
\addcomment{A replication factor of 5 incurs substantially higher memory, storage, and network cost without delivering significant performance benefit, while a replication factor of 2 
is unsupported by the Raft protocol.}

\noindent\textbf{Distributed Data Store.}
{\proj} offloads large object (e.g., model parameters) storage and replication to a \addcomment{pluggable} Distributed Data Store (Step~\circled{7}). These objects are asynchronously replicated when {\proj} detects GPU overload and switches to a new executor replica or migrates the current one to a different server (\cref{subsubsec:state_synchronization}). \addcomment{{\proj} supports Redis~\cite{redis}, AWS S3~\cite{aws_s3}, and HDFS~\cite{hdfs_msst10, hopsfs_fast17, lambdafs_asplos23}.}

\vspace{-2pt}
\subsection{Distributed Notebook Kernels }
\label{subsec:kernel_creation}

\subsubsection{Distributed Kernel Creation}
\label{subsubsec:kernel_creation}

\begin{figure}[t]
\centering
\includegraphics[width=0.4\textwidth]{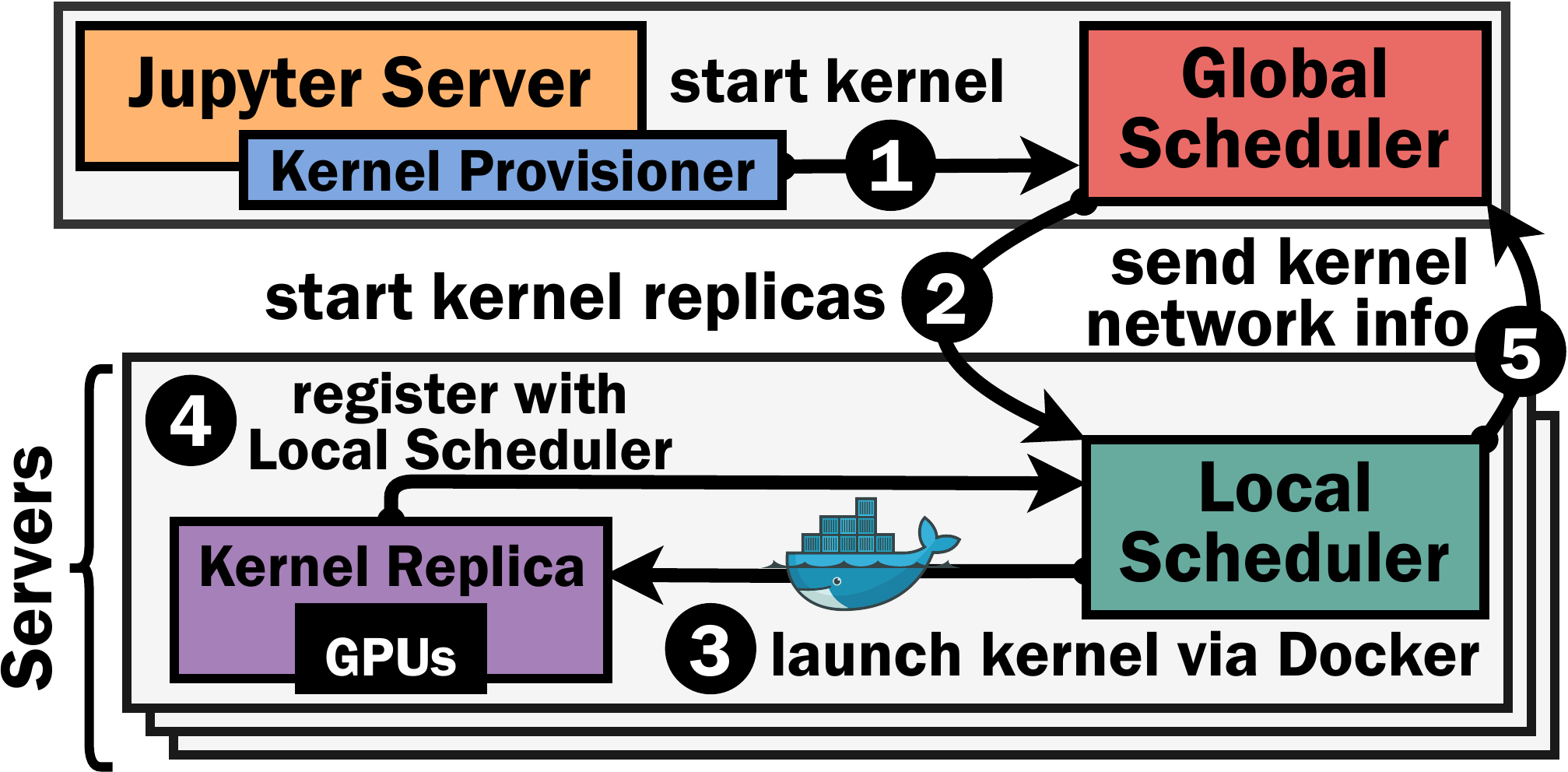}
\vspace{-6pt}
\caption{Process of creating a new kernel within {\proj}'s Docker Compose and Docker Swarm modes.}
\Description[Process of creating a new kernel within {\proj}'s Docker Compose and Docker Swarm modes.]{Process of creating a new kernel within {\proj}'s Docker Compose and Docker Swarm modes.}
\label{fig:create_kernel_diagram}
\vspace{-14pt}
\end{figure}

As described in \cref{subsec:design_overview}, {\proj} provides a customized Kernel Provisioner to integrate {\proj} directly with Jupyter. When a user creates a new Distributed Kernel, the Jupyter Server will create a new instance of {\proj}'s Kernel Provisioner, which issues a {\texttt{\small{StartKernel}}} RPC request to the Global Scheduler (see Step~\circled{1} in Figure~\ref{fig:create_kernel_diagram}).

Upon receiving the {\texttt{\small{StartKernel}}} RPC request, the Global Scheduler identifies three candidate GPU servers to host the replicas of the new kernel and then issues a {\texttt{\small{StartKernelReplica}}} RPC request to the Local Scheduler running on each of the candidate servers (Step~\circled{2}). 
The Global Scheduler eventually receives connection information about the new cluster of kernel replicas from the associated Local Schedulers and returns this information back to the Jupyter Server, which indicates that the new kernel was successfully created.

The Global Scheduler begins the process of identifying viable candidate GPU servers by examining the \emph{resource request} argument of the {\texttt{\small{StartKernelReplica}}} RPC. The \emph{resource request}, configured by the user, specifies the required resources for its notebook IDLT tasks, including CPUs (in millicpus, where 1 millicpu is equal to 1/1000$^{th}$ of a vCPU), memory (in megabytes), GPUs, and VRAM (in gigabytes). 

The Global Scheduler iterates over the servers in the cluster, checking if each server has sufficient capacity to serve a replica of the kernel. If a server meets the requirements, this server becomes a candidate.
When more than three viable candidates ($N > 3$) are found, a pluggable policy is used to select the target hosts. By default, the \textit{least-loaded} hosts---those with the fewest actively used GPUs---are chosen.

Note that, when scheduling a new Distributed Kernel, resources are not \emph{exclusively committed} to its replicas. Instead, the kernel replicas ``subscribe'' to the requested resources (Figure~\ref{fig:prototype_architecture}). 
Each server in the {\proj} cluster maintains a \emph{subscription ratio} with a configurable high watermark that prevents excessive over-subscription. 
This mechanism is designed to minimize resource contention among replicas of different Distributed Kernels scheduled on the same server. 

Upon receiving a {\texttt{\small{StartKernelReplica}}} RPC request, the Local Scheduler will provision a new container for the kernel replica (Step~\circled{3}), 
In Step~\circled{4}, the provisioned replica begins running by initiating a registration procedure with its Local Scheduler. During this procedure, the Local Scheduler informs the Global Scheduler that the replica has been created successfully. Finally, in Step~\circled{5}, the Local Scheduler concludes this process by returning the replica's connection information to the Global Scheduler.

After registering with their Local Schedulers, the three kernel replica containers establish peer-to-peer (P2P) connections with one another to run the Raft protocol for transparent kernel SMR. 
Once the Raft cluster is established, the kernels notify their Local Schedulers, which %
notify the Global Scheduler. The Global Scheduler then returns from its {\texttt{\small{StartKernel}}} RPC handler, and the kernel is officially created. 

If the Global Scheduler cannot identify three candidate servers to host the replicas of the new Distributed Kernel, then it will invoke a pluggable handler that is set based upon the configured scheduling policy. This invocation will initiate a scale-out operation to provision however many additional servers are required, as detailed in \cref{subsubsec:scale_out_operations}.

\subsubsection{Distributed Kernel Raft Protocol}
\label{subsubsec:kernel_raft_protocol}

\begin{figure}[t]
\centering
\includegraphics[width=0.5\textwidth]{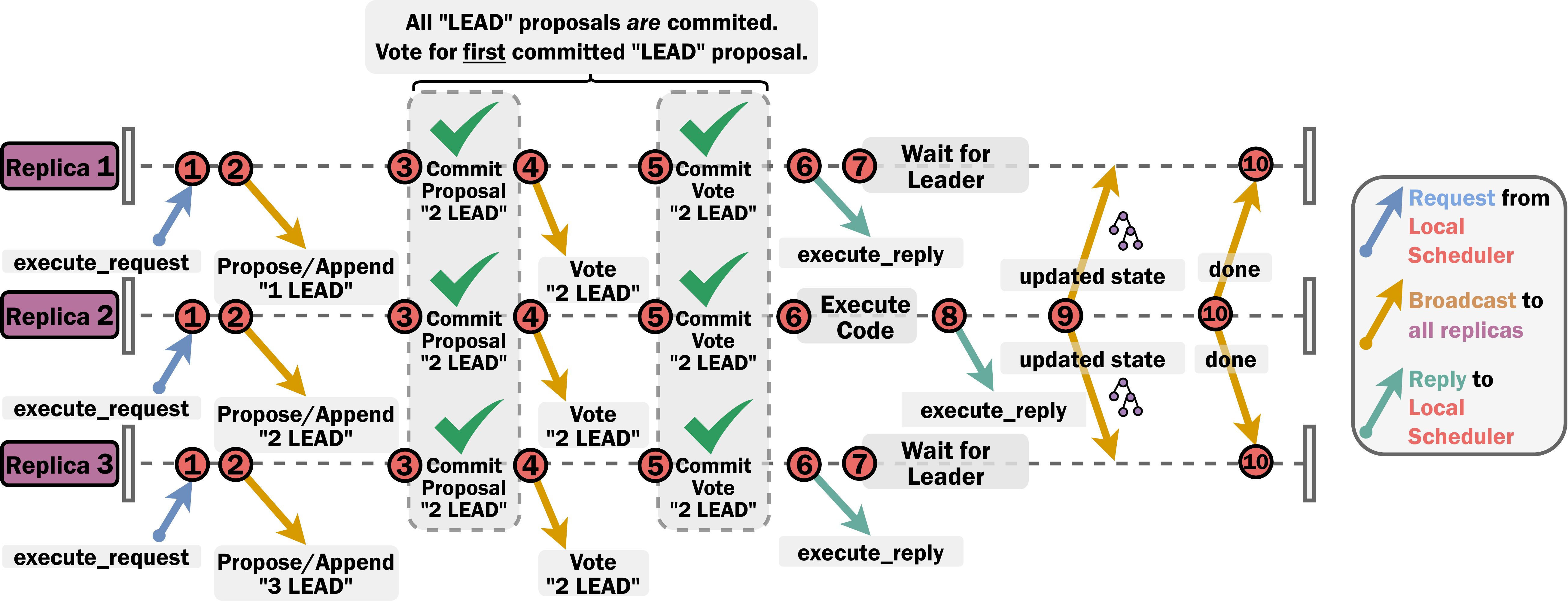}
\vspace{-14pt}
\caption{{\proj}'s executor election protocol.} 
\Description[{\proj}'s executor election protocol.]{{\proj}'s executor election protocol.}
\label{fig:kernel_raft_protocol}
\vspace{-14pt}
\end{figure}

The three replicas of a {\proj} Distributed Kernel connect to one another directly, forming a peer-to-peer (P2P) network. The replicas use Raft~\cite{raft_atc14} to perform SMR. The use of Raft is motivated by the need for high-availability and durable sessions, even for CPU-only notebooks: existing platforms like Google Colab reclaim idle sessions~\cite{colab_faq}, resulting in the loss of computed state. 
\addcomment{While recent progress on GPU state checkpoint/restore offers efficient mechanisms for GPU state snapshotting~\cite{phoenixos_sosp25, modal_gpu_snapshot}, coarse-grained CPU state checkpoint/restore~\cite{criu} is often unnecessarily heavyweight. {\proj} simplifies fault tolerance, replica migration, failover, and the prevention of duplicate cell execution by leveraging the strong consistency guarantees provided by Raft. 
To this end, {\proj} adopts a hybrid approach: it performs online replication for small CPU state and uses a distributed data store for asynchronous \diffcomment{persistence}{checkpointing and replication} of large objects.} 

Kernel replicas perform an \textit{executor replica election protocol} to designate an ``executor replica'' each time a user executes a notebook cell. This protocol is designed to be efficient and fault tolerant: progress occurs even when messages between replicas---or from each replica's respective Local Scheduler---are dropped or delayed. %
The %
executor replica is responsible for executing user-submitted code while the other replicas %
remain idle until the next cell execution request.

Figure~\ref{fig:kernel_raft_protocol} provides an overview of the executor election protocol. 
When a user submits a cell request, the client sends an {\texttt{\small{execute\_request}}} message to the Jupyter Server. This message, which encodes the target kernel's unique ID, is then forwarded to the Global Scheduler. The Global Scheduler then broadcasts a copy of this message to each Local Scheduler managing a server where a kernel replica container is running 
(see Step~\circled{1} in Figure~\ref{fig:kernel_raft_protocol}). 
When the Global Scheduler has sufficient resource information, it directly selects the executor replica for a given kernel and bypasses the Raft-based {\small\texttt{LEAD}}/{\small\texttt{YIELD}} election phase entirely. In this case, it will convert the {\texttt{\small{execute\_request}}} message into a {\texttt{\small{yield\_request}}}, signaling to the recipient kernel replicas that they should not attempt to participate in the election process and instead defer execution to the designated replica. This message conversion typically occurs when the Global or Local Scheduler determines that a particular server lacks the necessary resources for that server's kernel replica to execute code.

Upon receiving an {\texttt{\small{execute}}} or {\texttt{\small{yield}}} message, a kernel immediately appends a {\texttt{\small{LEAD}}} or {\texttt{\small{YIELD}}} proposal to its Raft log (Step~\circled{2}).  
In Step~\circled{3}, the kernel replicas wait for all  Step~\circled{2}'s proposals to be committed by Raft to the Raft log.  
Proposal commitment is handled by the Raft protocol. 
Kernel replicas will simply take note of any {\texttt{\small{YIELD}}} proposals that are committed, as the election will ``fail'' if all kernel replicas propose {\texttt{\small{YIELD}}}. This case is described in \cref{subsubsec:kernel_raft_protocol_failed_elections}. 

Once the first {\texttt{\small{LEAD}}} proposal is committed by Raft---for example, in Figure~\ref{fig:kernel_raft_protocol}, after Step~\circled{2}, Raft commits ``{\small\texttt{2 LEAD}}'' before ``{\small\texttt{1 LEAD}}'' and ``{\small\texttt{3 LEAD}}'' (Step~\circled{3})---the kernel replicas each vote for {\small\texttt{Replica 2}} by appending a {\small\texttt{VOTE}} proposal to the log  (Step~\circled{4}). 
The replicas encode the ID of the replica ({\small\texttt{Replica 2}}) that proposed the {\texttt{\small{LEAD}}} proposal in the {\texttt{\small{VOTE}}} proposal. This ID is used by the replicas in Step~\circled{5} to determine if they won the election.  
The kernel replica who won the executor election becomes the \emph{executor replica} and proceeds to execute the user-submitted code, while the other replicas become the \emph{standby replicas} and remain idle (Step~\circled{6}).  

Once the executor replica finishes executing the cell task, it commits a notification to the Raft log to inform its peers that the execution has finished (Step~\circled{7}). This notification is thereby received in Step~\circled{8}. In Step~\circled{9}, all replicas send a Jupyter {\texttt{\small{execute\_reply}}} message to their Local Scheduler, which forwards the message to the Global Scheduler. These messages are aggregated and merged together by the Global Scheduler before being forwarded back to the Jupyter Server and subsequently the user's client.  

After responding to the user's code submission, the executor replica begins replicating any updated notebook state to the standby replicas. Small state objects are directly replicated using {\proj}'s Raft SMR protocol. 
Large objects (model parameters copied from GPU VRAM and training datasets) are asynchronously written to the Distributed Data Store. This process occurs entirely outside the user request's critical path, avoiding any impact on user experience (\cref{subsubsec:state_synchronization}).

\vspace{-6pt}
\subsubsection{Handling Failed Executor Elections}
\label{subsubsec:kernel_raft_protocol_failed_elections}

In the worst case, if all kernel replicas {\texttt{\small{YIELD}}}, the Global Scheduler initiates a \textit{migration} of one of the kernel replicas to a server with sufficient resources. 
Specifically, the Global Scheduler selects a particular kernel replica for migration and instructs that replica to persist any important state to the Distributed Data Store. The kernel replica notifies the Global Scheduler once it has done so and is ready to migrate.  

Meanwhile, the Global Scheduler selects a target server as the destination of the replica's migration. The selection criteria depends on the cluster's configured scheduling policy. In general, the target server must have sufficient idle resources to immediately and exclusively bind the required GPUs to the migrated kernel replica. If a suitable server is found, the Global Scheduler sends a {\texttt{\small{StartKernelReplica}}} RPC to the Local Scheduler on that server, instructing it to provision a new kernel replica container. If no viable servers are available, the migration is enqueued and  periodically retried, several times if necessary, before ultimately being aborted if unsuccessful. In case of an aborted migration, an {\texttt{\small{execute\_reply}}} message with an error is returned to the client.

Once the new kernel replica has started and has read the persisted state from \diffcomment{remote storage}{the Distributed Data Store}, the Global Scheduler terminates the original kernel replica \diffcomment{before instructing}{. The Global Scheduler then instructs} the remaining replicas to reconfigure their Raft cluster to replace the terminated kernel \diffcomment{replica}{replica's ``peer address''}
with %
the newly-created replica. Next, the new replica joins its Raft cluster and begins replaying its Raft log \diffcomment{synchronize its state with its peers}{to restore its state to that of its peers}, after which the Raft cluster %
becomes operational again. %
Finally, the Global Scheduler resubmits the \diffcomment{execution request}{original {\texttt{\small{execute\_request}}}} to the migrated kernel replica, ensuring that it \diffcomment{executes}{is the one to execute} the user-submitted code while the other replicas yield.

\noindent\textbf{Pre-warmed Container Pool.} 
To reduce migration overhead, the Global Scheduler maintains a small pool of \emph{pre-warmed containers}~\cite{serverless_in_the_wild_atc20, faascache_asplos21}, managed by a component called the \emph{Container Prewarmer}. 
The Container Prewarmer uses a pluggable policy for provisioning an initial pool of warm containers, and another pluggable policy for maintaining the capacity of this pool of warm containers. By default, the Container Prewarmer ensures that each server has a specified, minimum number of pre-warmed containers available.

During kernel replica migrations, the Global Scheduler will query the Container Prewarmer to see if the selected target host has any pre-warmed containers available. If so, then a pre-warmed container will be used. These containers contain a pre-initialized Python runtime with commonly used dependencies, thereby eliminating the on-demand container provisioning overhead. Pre-warming containers with varied runtime dependencies is beyond the scope of this work; however, {\proj} could leverage existing checkpoint/restore~\cite{sock_atc18} or fork-based~\cite{vhive_asplos21} solutions for this purpose.

\subsubsection{Kernel State Replication} 
\label{subsubsec:state_synchronization}

\begin{figure}[t]
\centering
\includegraphics[width=0.42\textwidth]{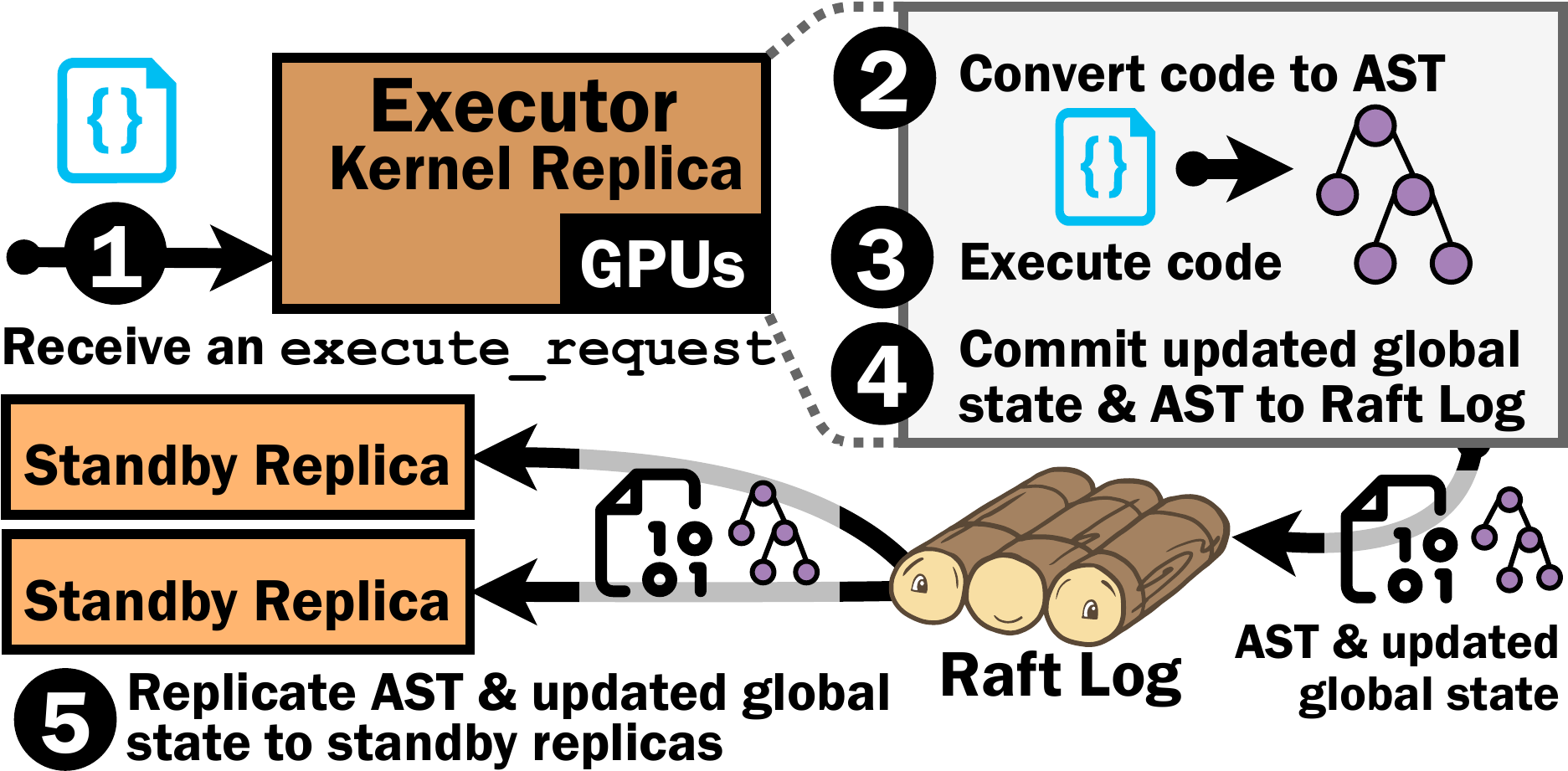}
\vspace{-8pt} 
\caption{Overview of the Raft-based state synchronization protocol used by {\proj}'s Distributed Kernel.} 
\Description[Overview of the Raft-based state synchronization protocol used by {\proj}'s Distributed Kernel.]{Overview of the Raft-based state synchronization protocol used by {\proj}'s Distributed Kernel.}
\label{fig:kernel_state_synchronization}
\vspace{-14pt} 
\end{figure}

{\proj}'s Distributed Kernel requires an efficient mechanism to maintain consistency across the kernel replicas. To achieve this, the kernels leverages Python AST-based code analysis~\cite{python_ast} to identify runtime state that needs to be replicated and synchronized via the Raft SMR. This procedure can detect and replicate Python-level state as well as state declared in native (C/C++) code, as this state is referenced in the kernel namespace (i.e., the set of user-defined variables stored within the memory of a kernel process). State of external processes or {\texttt{\small{libC}}} cannot be synchronized under the current implementation and is left as future work.
An overview of this protocol is shown in Figure~\ref{fig:kernel_state_synchronization}. As shown in Steps~\circled{1} and \circled{2}, code submitted for execution is first transformed into an AST by the executor replica before being executed in Step~\circled{3}. The executor replica then analyzes the Python AST to identify variables that must be synchronized with the executor's peers, \added{such as global variables storing intermediate state that are likely to be referenced later when executing subsequent notebook cells.} After the AST analysis, the executor replica proposes the AST and any relevant state to the Raft log in Step~\circled{4}. The Raft consensus protocol 
replicates the update to the distributed log across all replicas. Upon commitment of the newly appended log entry in Step~\circled{5}, the AST is applied by the standby replicas, and any changes to standby replicas' global state are applied to the replica-local variables.

\noindent\textbf{Handling Large Objects.} 
During state replication, the executor replica analyzes variable types from AST inspection. For large variables like models and datasets (hundreds of MBs to GBs), it avoids direct Raft log replication. Instead, it appends a \textit{pointer} to the log while storing the actual data in the Distributed Data Store (see Step~\circled{7} in Figure~\ref{fig:prototype_architecture}).

{\proj} supports AWS S3~\cite{aws_s3}, HDFS~\cite{hdfs_msst10}, and Redis~\cite{redis}. Pointers in the Raft log encode data retrieval, while kernel replicas handle persistence and retrieval transparently. If a user submits an {\texttt{\small{execute\_request}}} during state replication, it is enqueued until replication completes.

Large object replication occurs asynchronously between kernel replicas, but high task IAT in IDLT workloads hides this latency from users. {\proj} also employs a simple node-level cache to limit storage and memory costs. 

\vspace{-6pt}
\subsubsection{Handling Failures}
\label{subsubsec:design_handling_failures}

Each Distributed Kernel cluster is able to tolerate a \emph{fail-stop}~\cite{failstop_tocs83} failure of a single replica, as the Raft cluster of Distributed Kernel has 3 servers. If two or more replicas of a kernel were to fail, then this failure would be detected by the Global and Local Schedulers. 
{\proj}'s Distributed Kernel uses \emph{heartbeat messages}
to ensure that all components are active. If a heartbeat or another message, such as an {\texttt{\small{execute\_request}}}, times out, the Global Scheduler deems that a kernel has failed. In this case, the kernel's replicas can be terminated and recreated, and the replica's can restore all state from remote storage.

\vspace{-2pt}
\subsection{GPU Management}
\label{subsec:gpu_management}

{\proj} performs dynamic GPU binding right before a replica begins executing user-submitted code so as to optimize resource management by enabling fine-grained, per-training-task allocations (see step 6 in Figure~\ref{fig:prototype_architecture}).  
{\proj}'s approach to dynamic GPU sharing is also designed to be \emph{pluggable}, allowing integration with alternative approaches such as GaiaGPU~\cite{gaia_gpu}. 
We implement a simple approach in {\proj}'s prototype, with the intent of using a full-featured solution in production.

{\proj} binds all GPUs available on a server to all its hosted kernel replica containers. Each time a notebook cell is submitted for execution by a client, the Global Scheduler embeds 
the device IDs of the GPUs allocated to the target kernel replica within the request metadata. %
{\proj} automatically loads model parameters from the host's main memory onto all allocated GPUs using the {\texttt{\small{PyTorch}}} API on the critical path of execution requests. This process typically only takes up to a couple hundred milliseconds, so it does not impact performance or interactivity severely. When the executor replica finishes executing the user-submitted code, {\proj} automatically copies the user's data from the GPUs to the server's host memory. 
The executor replica returns the result to the user only after the GPU operations finish and GPU state is copied to host memory.

\vspace{-2pt}
\subsection{Resource Scheduling}
\label{subsec:resource_scheduling}

\subsubsection{{\proj}'s Default Placement Policy}
\label{subsec:default_placement_policy}

{\proj} is designed to be highly modular. The system can support arbitrary resource scheduling policies, and implementing support for a new policy is accomplished by implementing a simple interface. {\proj}'s default kernel replica  placement policy takes several different factors into account. \emph{First}, {\proj} considers the number of idle GPUs available on each GPU server. In particular, {\proj} favors placing kernel replica containers on servers with more idle GPUs available.

\emph{Second}, {\proj} considers a metric referred to as the \textit{subscription ratio} (SR) of a GPU server. The SR of a GPU server is defined as $\frac{S}{G \cdot R}$, where $S$ stands for the number of \emph{subscribed} GPUs, which is the sum of all the GPUs requested
by kernel replicas scheduled on that server (including idle kernel replicas), $G$ is the number of server GPUs (ranging from 1 to 8), and $R$ is the number of replicas per distributed kernel ($R=3$). $R$ accounts for the fact that, at a given time, only one out of $R$ replicas per distributed kernel will serve as the executor and actively use GPUs. Dividing by $R$ adjusts for the redundancy introduced by {\proj}'s distributed kernel, ensuring that the SR captures the \emph{effective} GPU subscription ratio rather than the total GPU subscription capacity of a single server. 
For example, if server $H$ with 8 GPUs is serving 4 kernel containers each requiring 4 GPUs,
then the number of subscribed GPUs $S$ for $H$ is $4\times4=16$. Consequently, $H$'s SR is $\frac{16}{8 \cdot 3}=0.667$. A cluster-wide SR of 1 or lower theoretically ensures that, during GPU code execution, any kernel replica $k$ can acquire its required number of GPUs on a server that is hosting $k$, where $k$ is one of the $R$ replicas of some distributed kernel.

\emph{Third}, {\proj} implements a dynamic, cluster-wide limit on the SR. While the cluster size is a function of the number of GPUs %
used by actively-training kernel replicas (detailed in \cref{subsubsec:scale_out_operations}), the maximum permissible SR across the entire cluster is dynamically adjusted.  This limit is calculated as $\frac{\sum{S}}{\sum{G} \cdot R}$, where $\sum{S}$ denotes the total number of subscribed GPUs for all kernel replicas across all servers, and $\sum{G}$ represents the total GPU count across all servers. If scheduling an additional kernel replica on a server would cause the server's SR to exceed this limit, the server is rejected in favor of another. The effect of this cluster-wide SR limit is illustrated in the evaluation results discussed in \cref{subsubsec:eval_prototype_subscription_ratio}.

\vspace{-4pt}
\subsubsection{Scale-Out Operations}
\label{subsubsec:scale_out_operations}

Scale-out operations involve provisioning additional servers in a platform-dependent manner, and then waiting for the Local Schedulers that are started on the new servers to connect and register with the Global Scheduler. Scale-out operations occur in one of two scenarios. \emph{First}, they are triggered in response to a failed attempt to place one or more replicas of a distributed kernel. Such a failure occurs when there are no viable candidate servers across the cluster to serve the kernel replicas. Upon triggering the scale-out operation, the placement of the corresponding kernel replicas is paused. Resources are immediately reserved for the paused kernel replicas on newly provisioned servers, i.e., before the servers are fully added to the {\proj} server cluster. Once the servers are ready, %
{\proj} resumes placing the replicas on them. 

\emph{Second}, scale-out operations can also be triggered by {\proj}'s auto-scaling policy. {\proj}'s auto-scaler runs on a configurable interval, monitoring cluster resource utilization and determining whether servers should be added or removed. To handle bursts of training requests, {\proj} maintains a \emph{scaling buffer} of ``extra'' servers. 
To decide if additional servers should be added, the auto-scaler first examines the total number of GPUs actively committed ($\sum C$) to kernel replicas (that are actively executing GPU code) on servers across the {\proj} cluster, where $C$ is the number of actively-utilized GPUs on a server. The expected cluster capacity ($\sum G^\prime$) is defined as $\sum G^\prime = f\cdot \sum C$, where $f$ is a scalar multiplier that controls how aggressively {\proj} scales. If the current cluster capacity is smaller than $\sum G^\prime$, additional GPU servers are provisioned. 
Intuitively, if the number of actively-utilized GPUs is low, then the auto-scaler will not try to add additional servers. If the number of actively-utilized is GPUs is high and many existing servers are at capacity, then {\proj}'s auto-scaler triggers scale-out. 
We set $f$ to 1.05, as we have found empirically that this enables an appropriate degree of auto-scaling during testing. 

If cluster resource usage is too low, {\proj}'s auto-scaler attempts to release 1-2 idle servers at a time (where idle servers are those with no active training kernel replicas.) The auto-scaler determines if scaling-in is appropriate using a similar approach to scaling-out%
: if $\sum G^\prime$ is less than the number of currently-provisioned GPU servers within the cluster, then
{\proj} gradually releases servers 1 to 2 at a time until this condition is no longer met.

\vspace{-2pt}
\section{{\proj} Implementation}
\label{sec:implementation}

To demonstrate {\proj}'s efficacy, we implemented a fully functional prototype Jupyter Notebook platform.
{\proj} maintains compatibility with all Jupyter clients by reusing the IPython messaging protocol~\cite{ipython_messaging_protocol}. It leverages Jupyter Server from the official \texttt{\small{base-notebook}} Docker image~\cite{base_notebook_docker} and introduces custom components for {\proj}-specific functionality. All custom components were implemented using the official Jupyter Server API extension methods~\cite{jupyter_kernel_provisioning}, building on the default Jupyter Server.

{\proj} is 
platform-agnostic. We developed and evaluated {\proj} atop Docker Swarm~\cite{docker_swarm} and Docker Compose~\cite{docker_compose}; Kubernetes~\cite{kubernetes} is also supported. To simplify deployment, experimentation, and reproducibility, we provide a set of Ansible~\cite{ansible} playbooks to automate setup, installation, and deployment on these container orchestration platforms. 
We have implemented {\proj} 
in approximately 282k lines of code  over roughly two person-years. 
Refer to Appendix~\ref{sec:appendix-implementation} for additional implementation details. 

\vspace{-2pt}
\section{Evaluation}
\label{sec:evaluation}

In this section, we present our evaluation of {\proj}.

\vspace{-2pt}
\subsection{Experimental Setup \& Methodology}
\label{subsec:eval_setup_methodology}

\subsubsection{Baselines} 
\label{subsubsec:eval_setup_methodology_baselines}

We implement three representative baseline policies within {\proj} itself, each of which represents a unique category of alternative solutions. %
 
\noindent\textbf{{\reservation}}
emulates the behavior of current notebook platforms like {\idltplatform} and Google Colab. {\reservation} creates one long-running kernel container for each user session that remains active for the entire duration of the user session. Fixed resources, including GPUs, are exclusively allocated to each long-running kernel container. 

\noindent\textbf{\FCFS}
is a baseline representing the class of batch GPU cluster schedulers~\cite{gandiva_osdi18, tiresias_nsdi19, themis_nsdi20, pollux_osdi21, gavel_osdi20} designed for long-running GPU training workloads. {\FCFS} provisions a kernel replica container each time a user submits code and a job request for execution (e.g., to a slurm scheduler). The new container serves the training request before terminating. As such, {\FCFS} approximates on-demand resource scaling. While various GPU schedulers differ in their design, implementation, and scheduling/placement algorithms, they all share several significant sources of overhead in the context of \emph{notebook scheduling} for IDLT workloads. We implemented {\FCFS} using a first-come, first-serve (FCFS) job scheduling and GPU allocation policy within {\proj} to approximate the performance of these GPU schedulers.  

\noindent\textbf{{\projlcp}} 
(large container pool) is an alternative {\proj} baseline that sacrifices (some) interactivity in favor of reduced resource cost. {\projlcp} serves as a tool for exploring the trade-off between interactivity and resource efficiency. 
{\projlcp} employs a larger pool of pre-warmed containers than {\proj}'s default configuration, which uses three kernel replicas 
and a significantly smaller pre-warmed container pool used during  replica migrations. 
When a cell task arrives, {\proj} selects a warm container\footnote{We analyzed the Python dependencies of  notebook containers from our workload and observed substantial overlaps in common dependencies.} from the pool to serve the request. After execution, the container is returned to the pool rather than being terminated, as in {\proj}'s default policy. Addressing security and privacy concerns related to container sharing is out of scope for this paper but can leverage existing research~\cite{trenv_sosp24, rainbowcake_asplos24, pagurus_atc22}.

\subsubsection{Setup}
\label{subsubsec:eval_setup}

\begin{table}[t]
\caption{Models and datasets used in the evaluation along with their associated application domains.} 
\vspace{-8pt}
\label{tab:datasets_and_models}
\scalebox{0.875}{ %
\begin{tabular}{@{}lll@{}}
\toprule
{\bf App domain}                                                                  & {\bf Dataset}                                                                        & {\bf Model}                                                                        \\ \midrule
\begin{tabular}[c]{@{}l@{}}Computer\\ Vision\end{tabular}             & \begin{tabular}[c]{@{}l@{}}CIFAR-10,\\ CIFAR-100, \\ Tiny ImageNet\end{tabular} & \begin{tabular}[c]{@{}l@{}}VGG-16,\\ ResNet-18, \\ Inception v3\end{tabular} \\ \midrule
\begin{tabular}[c]{@{}l@{}}Natural Language\\ Processing\end{tabular} & \begin{tabular}[c]{@{}l@{}}IMDb Large Movie\\ Reviews, CoLA\end{tabular}        & BERT, GPT-2                                                                  \\ \midrule
\begin{tabular}[c]{@{}l@{}}Speech\\ Recognition\end{tabular}          & LibriSpeech                                                                     & Deep Speech 2                                                               
\end{tabular}
}
\end{table}

We conducted our evaluation on AWS EC2~\cite{ec2}. We ran each baseline on a cluster of 30 GPU EC2 VMs, each equipped with 8 GPUs to match the {\idltplatform} production setups. 
We developed a workload driver and dashboard to automate workload deployment and execution on {\proj}.
Following~\cite{elasticflow_asplos23}, we integrated support for a diverse set of DL models and datasets across several different application domains (Table~\ref{tab:datasets_and_models}). %
The workload driver randomly assigns each client an application domain, after which a random dataset and model are assigned. Each cell task request submitted by the client will train its assigned model on the assigned dataset, which are retrieved by kernel replicas from an AWS S3 bucket. %

We evaluate {\proj} using a 17.5-hour excerpt of the {\idltplatform} workload trace. To further elucidate its performance and monetary cost implications of {\proj}, we also implemented a robust, detailed simulator that we used to run the full {\idltplatform} trace. 
The simulator implemented {\proj}'s default scheduling/placement policy along with the baseline policies described in \cref{subsubsec:eval_setup_methodology_baselines}. The results of our simulation study on {\proj} are presented in \cref{subsec:eval_sim_study}.

\vspace{-2pt}
\subsection{Prototype Evaluation}
\label{subsec:eval_prototype}

This section presents the results of our prototype evaluation.

\vspace{-2pt} 
\subsection{Active Sessions \& Training Events}

\begin{figure}[t]
\centering
\includegraphics[width=0.45\textwidth]{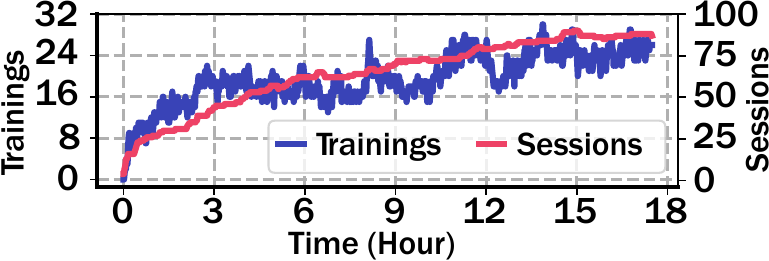}
\vspace{-10pt} 
\caption{The number of active user-submitted training tasks and active user sessions during the 17.5-hour {\platformtrace}.} 
\Description[The number of active user-submitted training tasks and active user sessions during the 17.5-hour {\platformtrace}.]{The number of active user-submitted training tasks and active user sessions during the 17.5-hour {\platformtrace}.}
\label{fig:active_sessions_and_trainings_18_hours}
\vspace{-12pt} 
\end{figure}

\begin{figure*}[t]
\begin{center}
\includegraphics[width=1.0\textwidth]{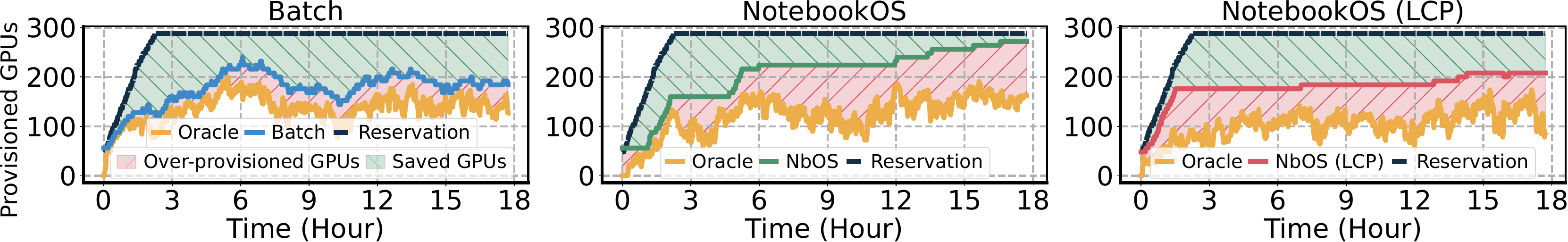}
\vspace{-17pt}
\caption{Provisioned GPUs timelines. Curves for {\FCFS} (left), {\proj} (middle), and {\projlcp} (right) represent the number of GPUs provisioned under each policy, while ``oracle'' represents an optimal policy that provisions the exact number of GPUs required to serve training requests. The %
GPUs saved relative to {\reservation} is shown in the green-shaded region, while the number of GPUs over-provisioned compared to ``oracle'' are represented by the orange-shaded region.}
\Description[GPU usage timelines.]{GPU usage timelines. Curves for {\FCFS} (left), {\proj} (middle), and {\projlcp} (right) represent the number of GPUs provisioned under each policy, while ``oracle'' represents an optimal policy, which provisions the exact number of GPUs required to serve training requests. The number of GPUs saved relative to {\reservation} is shown in the green-shaded region, while the number of GPUs over-provisioned compared to ``oracle'' are represented by the orange-shaded region.}
\label{fig:eval-allocatable-gpus}
\vspace{-8pt}
\end{center}
\end{figure*}

Figure~\ref{fig:active_sessions_and_trainings_18_hours} presents a timeline plot showing the number of active sessions during the execution of the 17.5-hour workload trace excerpt on {\proj}. This series is plotted on the secondary (i.e., right) y-axis. The number of active sessions increases from 0 at the beginning of the trace excerpt to 87 by the end of the excerpt. The maximum number of active sessions at any given time throughout the 17.5-hour {\platformtrace} excerpt is 90.

Figure~\ref{fig:active_sessions_and_trainings_18_hours} also presents a timeline plot showing the number of user-submitted training events being processed during the execution of the 17.5-hour {\platformtrace}  excerpt on {\proj}. This series is plotted on the primary (i.e., left) y-axis. Initially there are no active, user-submitted trainings. At the end of the trace excerpt, there are 26 active user-submitted trainings. The mean and medium number of active, user-submitted trainings are 19.5 
and 19, respectively. The maximum number of active user-submitted trainings at any given time throughout the summer {\platformtrace}  is 141 (see Figure~\ref{fig:active_sessions_and_trainings_full_summer} in \cref{sec:appendix}). The maximum number of active user-submitted trainings at any given time throughout the 17.5-hour {\platformtrace} excerpt is 34.

\vspace{-3pt}
\subsubsection{Resource Usage \& Efficiency}
\label{subsubsec:eval_gpu_comparisons}

{\FCFS} (Figure~\ref{fig:eval-allocatable-gpus} left), achieves significantly improved resource utilization compared to {\reservation} because {\FCFS} only provisions containers and allocates resources in response to training requests. 
{\proj} achieves higher GPU utilization than {\reservation} (Figure~\ref{fig:eval-allocatable-gpus}, middle), saving 1,187.66 ({\proj}) and 1,662.53 ({\projlcp}) GPU hours. However, because {\proj} maintains three long-running replicas per kernel as well as a small buffer of ``extra'' GPU servers for request bursts, {\proj} provisions more servers than {\FCFS}, which only allocates GPUs during cell executions. {\projlcp} (Figure~\ref{fig:eval-allocatable-gpus}, right) improves resource efficiency, provisioning 23.52\% fewer GPUs than {\proj} but still 18.18\% more than {\FCFS}. Note that the oracle curve shows the number of GPUs required to serve all active training requests, rather than aggregated GPU bandwidth utilization, which is shown in Figure~\ref{fig:gpu_usage_timeline}. {\proj} provisions slightly more GPUs than Batch/LCP to account for its use of replicated kernels and to handle request bursts.

\begin{figure}[t]
\begin{center}
\subfigure[Interactivity delay.] {
\includegraphics[height=0.16\textwidth]{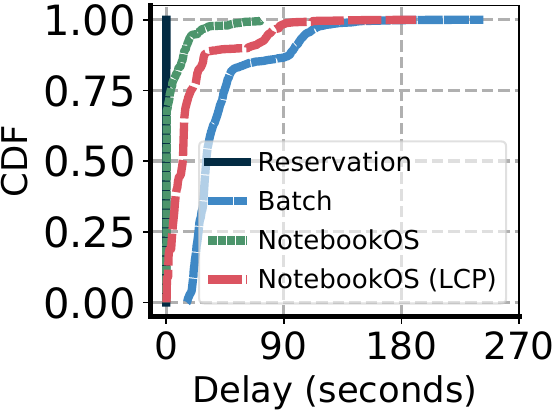}
\label{fig:eval_interactivity_delay}
}
\hspace{-4pt}
\subfigure[Task completion time (TCT).] {
\includegraphics[height=0.16\textwidth]{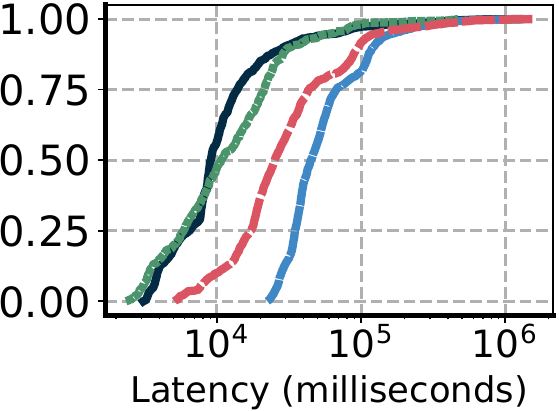}
\label{fig:eval_e2e_request_latency}
}
\vspace{-12pt}
\caption{CDFs of (a) \emph{interactivity delays} and (b) \emph{task completion times (TCT)} across different scheduling policies.} 
\Description[CDFs of (a) \emph{interactivity delays} and (b) \emph{task completion times (TCT)} across different scheduling policies.]{CDFs of (a) \emph{interactivity delays} and (b) \emph{task completion times (TCT)} across different scheduling policies.}
\label{fig:eval_gpu_usage_timeline}
\vspace{-10pt}
\end{center}
\end{figure}

\subsubsection{Interactivity}
\label{subsubsec:eval_prototype_interactivity}

Figure~\ref{fig:eval_interactivity_delay} plots a CDF of the per-task interactivity delays incurred by the different scheduling policies. The interactivity delay is the interval between the instant that a client submits a Jupyter ``{\texttt{\small{execute\_request}}}'' message to a kernel and the instant that the kernel begins executing the user-submitted code included within the 
message. 
Long interactivity delays are perceptible to clients and degrade the user experience. 

The {\reservation} baseline binds GPUs to a user's notebook kernel for its full lifetime, ensuring high interactivity. However, IDLT workloads often under-utilize GPU resources, as users frequently debug, edit code, or pause while GPUs remain idle but reserved. 

\noindent{\textbf{Effect of Multiple Kernel Replicas.}} 
{\proj} commits GPUs to a kernel replica immediately upon receiving an execution request 89.6\% of the time. Similarly, 89.45\% of the time, {\proj} reused the same executor replica for consecutive execution request. {\proj}'s use of replicated kernels maximizes the chances that at least one replica has available resources when a training request arrives, achieving nearly the same interactivity as the {\texttt{\small{reservation}}} baseline.

\subsubsection{Task Completion Time (TCT)}
\label{subsubsec:eval_task_completion_time}

TCT measures the interval between cell submission and the instant the cell's execution is completed. As shown in Figure~\ref{fig:eval_e2e_request_latency}, {\proj} achieves a TCT distribution comparable to {\reservation}, with slightly higher TCTs from the $38^{th}$ to the $90^{th}$ percentile.
This is because some sessions were not able to find available GPUs on their servers (due to oversubscription and server's full GPU usage) and thus are migrated to different servers. When {\proj}'s pre-warmed containers were exhausted, new containers were created on demand, incurring long cold startup delays. This can also be observed from the tail of {\proj}'s interactivity delays in Figure~\ref{fig:eval_interactivity_delay}. 
{\projlcp} sees much longer TCTs as a submitted cell request triggered a warming-up operation to download model parameters and datasets. 
FCFS yields the highest TCTs due to on-demand notebook kernel provisioning and mandatory pre- and post-processing data I/O (model parameters and datasets) for each submission.

\subsubsection{Subscription Ratio}
\label{subsubsec:eval_prototype_subscription_ratio}

\begin{figure}[t]
\centering
\includegraphics[width=0.44\textwidth]{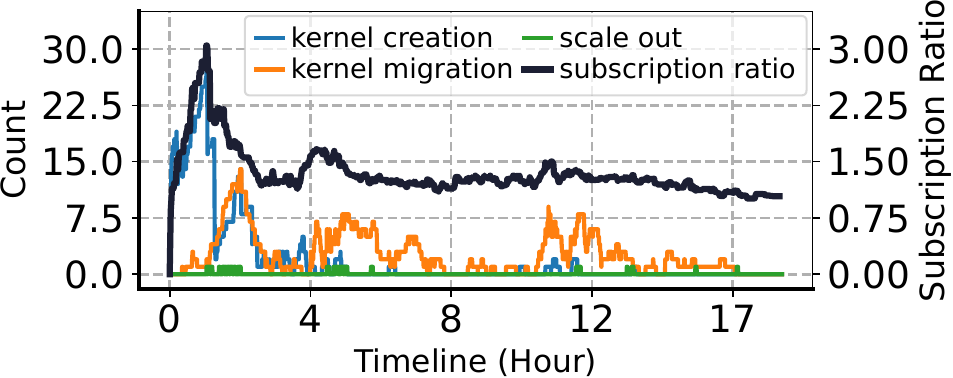}
\vspace{-10pt}
\caption{Timeline of major events occurred while executing the 17.5-hour workload on {\proj}, with {\proj}'s subscription ratio plotted on the secondary Y-axis.}
\Description[Timeline of major events occurred while executing the 17.5-hour workload on {\proj}, with {\proj}'s subscription ratio plotted on the secondary Y-axis.]{Timeline of major events occurred while executing the 17.5-hour workload on {\proj}, with {\proj}'s subscription ratio plotted on the secondary Y-axis.}
\label{fig:eval_subscription_ratio} 
\vspace{-8pt}
\end{figure}

The relationship between the cluster-wide subscription ratio (SR, described in \cref{subsec:resource_scheduling}) and the triggering of scale-out events by {\proj}'s auto-scaling policy (\cref{subsubsec:scale_out_operations}) are clearly shown in Figure~\ref{fig:eval_subscription_ratio}. Specifically, the SR increases sharply at the beginning of the workload when many new Distributed Kernels 
are created. 
To accommodate these kernel replicas, {\proj} triggers scale-out events. As new servers are provisioned, the cluster-wide SR falls. Around hour two, there is another burst of kernel replica creations, leading to another increase in the SR. Subsequently, additional servers are provisioned to serve the new kernel replicas shortly after hour four, leading to another drop in the SR. A similar pattern occurs before hour twelve. 
Similarly, there is an increase in the frequency of kernel migrations when the SR begins to climb, reflecting growing GPU resource saturation and potential contention when training requests arrive. Notably, a spike in migrations coincides with the first spike in the SR, followed by additional migrations after hours four and twelve of workload execution.

\subsection{Object Synchronization Overhead}
\label{subsec:object_synchronization_overhead}

\begin{figure}[t]
\centering
\includegraphics[width=0.45\textwidth]{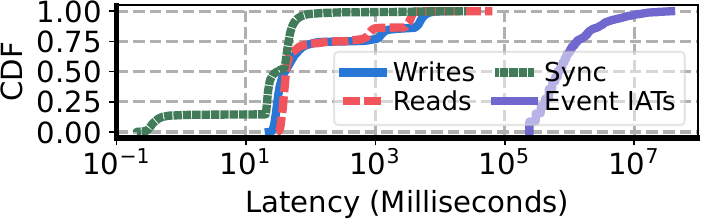}
\vspace{-10pt} 
\Description[CDF of the latency of reading and writing large objects and synchronizing smaller object state via the Raft synchronization protocol. Note that the x-axis is log-scale.]{CDF of the latency of reading and writing large objects and synchronizing smaller object state via the Raft synchronization protocol. Note that the x-axis is log-scale.} 
\caption{CDF of the latency of reading and writing large objects and synchronizing smaller object state via the Raft synchronization protocol. Note that the x-axis is log-scale.}
\label{fig:io_latency_cdf}
\vspace{-6pt} 
\end{figure}

Figure~\ref{fig:io_latency_cdf} presents a timeline plot showing the number of active sessions during the execution on {\proj} of the 17.5-hour {\platformtrace}  excerpt. %
The ``{\small\texttt{Sync}}'' curve corresponds to the end-to-end latency of synchronizing small objects via the Raft protocol (\cref{subsubsec:state_synchronization}). %
The ``{\small\texttt{Sync}}'' latency is typically extremely short: the $90^{th}$, $95^{th}$, and $99^{th}$ percentile values are 54.79ms, 66.69ms, and 268.25ms, respectively. The ``{\small\texttt{Reads}}'' and ``{\small\texttt{Writes}}'' curves correspond to the latencies observed when reading and writing large objects from and to the Distributed Data Store, respectively. These operations are reasonably short, as 99\% of {\texttt{\small{read}}} and {\texttt{\small{write}}} operations complete within roughly 3.95
and 7.07 
seconds, respectively.

As discussed in \cref{subsec:workload_analysis}, the $50^{th}$ and $75^{th}$ percentile of task IATs are 300 seconds (5 minutes) and 480 seconds (8 minutes) for the {\platformtrace}, respectively. 
The shortest event IAT within the {\platformtrace} is 240 seconds, or 4 minutes. Notably, the overhead of writing and reading large objects to and from the Distributed Data Store is completely contained within event IATs. As a result, this overhead is almost completely hidden from users in IDLT workloads such as the {\idltplatform} workload.

\subsection{Simulation Study}
\label{subsec:eval_sim_study}

This section present the results of our simulation study.

\vspace{-3pt}
\subsubsection{Monetary Cost}
\label{subsubsec:eval_sim_study_monetary_cost}

To understand the cost implications of {\proj},
we compare it with {\reservation} in terms of provider costs---specifically, the operational cost of provisioning EC2 resources, as done by production notebook platforms like {\idltplatform}---and the revenue generated by the notebook service provider. 

We implemented a simple billing model, assuming the provider covers AWS EC2 VM costs. Users pay 1.15× the provider's rate, proportional to resource usage. Standby Distributed Kernel replicas are charged 12.5\% of the base rate.
For example, if a provider pays \$10/hour for an 8-GPU EC2 VM, each standby replica is billed \$1.44/hour ($10 \times 1.15 \times 0.125$). When a replica runs a training task using 4 GPUs, the charge increases to \$5.75/hour ($10 \times 1.15 \times 0.5$). {\reservation} follows the same 1.15$\times$ multiplier for on-demand GPU usage.  

\begin{figure}[t]
\vspace{-5pt}
\begin{center}
\subfigure[Provider-side cost and revenue.] {
\includegraphics[height=0.1625\textwidth]{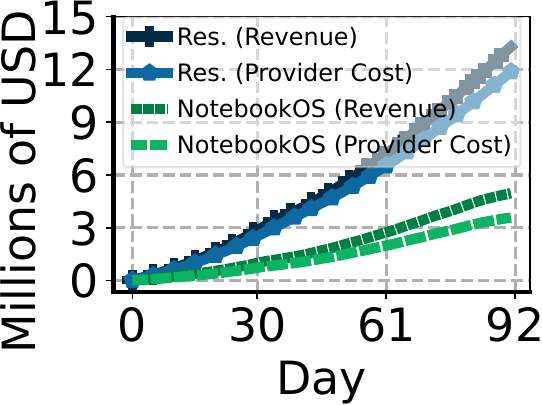}
\label{fig:eval_sim_overhead_revenue}
}
\hspace{-4pt}
\subfigure[Profit margin.] { 
\includegraphics[height=0.1625\textwidth]{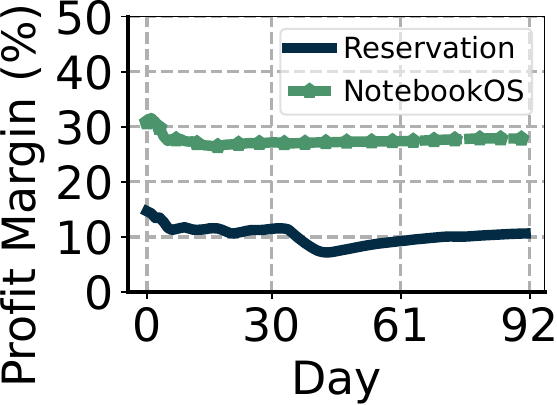}
\label{fig:eval_sim_profit_margin}
}

\vspace{-12pt}
\caption{Timeline of provider cost (provisioned EC2 resources), revenue, and profit margin.} 
\Description[Timeline of provider cost (provisioned EC2 resources), revenue, and profit margin.]{Timeline of provider cost (provisioned EC2 resources), revenue, and profit margin.}
\label{fig:eval_sim_overhead_revenue_profit_margin}
\vspace{-6pt}
\end{center}
\end{figure}

As shown in Figure~\ref{fig:eval_sim_overhead_revenue}, {\proj} achieves a provider-side \$-cost reduction of up to 69.87\% compared to {\reservation} by the end of the trace, thanks to its GPU resource savings. 
{\proj} also offer a higher profit margin (Figure~\ref{fig:eval_sim_profit_margin}) by significantly lowering provider costs and modestly increasing revenue through minimal charges on idle replicas. The increased profit comes with the added benefit of guaranteed programming interactivity.

\vspace{-2pt}
\subsubsection{GPU Hours Saved}
\label{subsubsec:eval_sim_study_resource_usage}

\begin{figure}[t]
\centering
\includegraphics[width=0.45\textwidth]{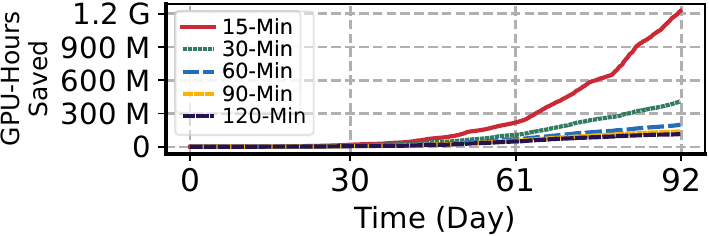}
\vspace{-10pt} 
\Description[The number of GPU hours saved by {\proj} by avoiding unnecessary re-execution of notebook cells following idle reclamations of user sessions.]{The number of GPU-hours saved by {\proj} by avoiding unnecessary re-execution of notebook cells following idle reclamations of user sessions.} 
\caption{The number of GPU-hours saved by {\proj} by avoiding unnecessary re-execution of notebook cells following idle reclamations of user sessions.} 
\label{fig:jupyter_gpu_hours_saved_trace}
\vspace{-10pt} 
\end{figure}

Figure~\ref{fig:jupyter_gpu_hours_saved_trace} illustrates the number of GPU hours saved by {\proj} by avoiding unnecessary re-execution of notebook cells following idle session reclamations across the entire {\platformtrace}. Each of the five lines in the plot represents a different configuration of the idle reclamation interval---the duration for which a Jupyter Notebook session may remain idle before its associated kernel is reclaimed. In the absence of {\proj}'s state replication and persistence mechanisms, reclaiming idle sessions would result in the loss of in-memory state, requiring cell re-execution and additional GPU usage when users return. By preserving session state, {\proj} prevents this redundant computation and significantly reduces GPU waste.

\vspace{-2pt}
\section{Related Work}
\label{sec:related_work}

\noindent\textbf{Notebook Platforms.}

Commercial notebook services such as Google Colab~\cite{colaboratory}, NaaS~\cite{naas}, Lentiq~\cite{lentiq}, and JupyterHub~\cite{jupyterhub} 
allow users to reserve GPUs for however long they like with great flexibility but low GPU utilization. These platforms
add keep-alive and timeout strategies 
so that the platform can automatically terminate a notebook instance if the user does not interact with it for a configurable period of time. The keep-alive strategy can help save unused GPU resources to some extent but does not fundamentally address the low GPU utilization issue. 
Azure Machine Learning~\cite{azure_ml}, Open OnDemand~\cite{open_ondemand, osc_jupyter_batchconnect}, among others~\cite{ncar_papermill, sagemaker_notebooks, sagemaker_studio_lab}, offer ``batch-connect'' notebook services to allow users to submit GPU jobs through a batch scheduler. 
This approach introduces significant runtime cold startup cost and batch scheduling delays during request bursts, undermining the interactive nature of notebook programming.

\noindent\textbf{GPU Cluster Schedulers.}
State-of-the-art batch GPU cluster schedulers~\cite{gandiva_osdi18, tiresias_nsdi19, themis_nsdi20, gavel_osdi20, pollux_osdi21} aim to balance the competing demands of long-running GPU training jobs but overlook responsiveness and interactivity, making them poorly suited for notebook IDLT workloads. 
Gandiva~\cite{gandiva_osdi18} is a GPU cluster scheduler that leverages domain-specific knowledge and an introspective scheduler to dynamically optimize BDLT workloads. 
Tiresias~\cite{tiresias_nsdi19} is a GPU cluster manager that minimizes job completion time %
by prioritizing jobs with partial information and tasks with the least service received. 
Themis~\cite{themis_nsdi20} employs a 2-tiered, ``finish-time fair'' approach where workloads bid for GPUs in an auction, balancing short-term efficiency with long-term fairness.

\noindent\textbf{Serverless.}
Optimized snapshot-loading reduces cold start costs for CPU~\cite{sock_atc18, faasnap_eurosys22, vhive_asplos21, aws_lambda_snapstart} and GPU~\cite{parallel_gpu_os, serverlessllm_osdi24, lambdascale_arxiv25} serverless functions.
Locality-driven keep-alive~\cite{peeking_atc18, faascache_asplos21, rainbowcake_asplos24, cidre_asplos25} optimize container caching for minimizing cold start costs.
{\proj}'s pre-warmed container pool can leverage these methods to further cut GPU startup cost.

\noindent\textbf{GPU Virtualization.}
Existing solutions for GPU virtualization and sharing leverage NVIDIA MIG~\cite{nvidia_mig} / MPS~\cite{nvidia_mps} as well as user-space CUDA API interposition and remoting techniques for GPU kernel performance isolation~\cite{tgs_nsdi23, tally_asplos25}, memory isolation~\cite{faaswap}, and parallel executions within a single remote GPU~\cite{rcuda_hpcs10}. Currently, {\proj} supports only coarse-grained, exclusive GPU time-sharing via kernel replication and executor replica election but does not support fine-grained, concurrent, spatial GPU sharing. Enabling fine-grained GPU sharing is part of our future work.

\noindent\textbf{Consensus Protocols and Leader Election Algorithms.}
Raft~\cite{raft_atc14} is a widely-adopted consensus algorithm designed for understandability 
while providing strong consistency, featuring leader elections with randomized timeouts (to prevent ties) and log replication via heartbeats. {\proj}'s state synchronization and executor selection protocols are implemented atop Raft, as they require more complex logic than what is provided directly by Raft's leader elections. The use of Raft provides robust, off-the-shelf consistency and fault tolerance support. 
Alternative consensus protocols, such as Viewstamped Replication~\cite{viewstamped_podc88}, EPaxos~\cite{epaxos_sosp13}, and Zab~\cite{zab_dsn11}, offer tradeoffs in latency, leader dependence, and throughput. 
The Bully algorithm~\cite{bully_tc82} targets crash-recovery systems and elects the node with the highest ID, while the Ring algorithm~\cite{ring_election_algo, ring_algo_lelann} organizes nodes into a logical ring in which communication occurs only between adjacent neighbors. 
Systems like ZooKeeper~\cite{zookeeper_atc10} and Chubby~\cite{chubby_osdi06} offer practical leader election mechanisms, informing the design of resilient control planes in {\proj}.

\vspace{-2pt}
\section{Limitations and Future Work}
\label{sec:discussion}

{\proj} does not yet support GPU sharing or fractional allocations. Future work will incorporate GPU virtualization using tools such as HAMi~\cite{hami}. 
While multi-server training is also not supported, it can be enabled through integration with distributed training frameworks like Ray~\cite{ray_osdi18}. 
These capabilities are largely orthogonal to {\proj}'s core abstractions, and {\proj}'s mechanisms for scheduling and execution would remain largely unchanged. \addcomment{Lastly, {\proj} could further optimize CPU-only notebook cells by enabling elastic burst-parallel executions through the integration of stateful serverless scheduling techniques~\cite{wukong_socc20, cloudburst_vldb20}.}

\vspace{-2pt}
\section{Conclusion}
\label{sec:conclusion}

The key insight of this paper is that notebooks are long-running but with fragmented executions, rendering intermittent, sporadic, and often low GPU utilization. 
We built {\proj}, a GPU-efficient notebook platform designed to meet the unique notebook workload requirements and prioritizes interactivity. {\proj} oversubscribes and multiplexes GPUs using a novel replicated kernel design. {\proj} dynamically allocates GPUs only for notebook cell executions that involve GPU training. We evaluated {\proj} using production notebook workloads. Results show that {\proj} reduces cluster GPU resource cost significantly compared to existing notebook platforms.
Integrating {\proj} in production 
is part of our future work.

\vspace{-2pt}
\section*{Acknowledgments} 
\label{sec:Acknowledgements}

We thank the anonymous reviewers and our shepherd, Rodrigo Bruno, for their valuable feedback and comments, which improved the paper.   
This research was supported in part by NSF grants OAC-2411009 and CNS-2322860 (NSF CAREER Award)\diffcomment{.}{,} 
We also acknowledge support from NSF CloudBank for providing AWS credits, and thank Adobe for their generous research gift. Benjamin Carver was supported by a Presidential Scholarship from George Mason University. 

\newpage 

\bibliographystyle{plain}
\balance
\bibliography{references}

\newpage

\clearpage
\appendix

\section{Artifact Appendix}
\label{sec:appendix}

\subsection{Abstract}

This section presents supplementary materials and guidance intended to support the reproducibility of {\proj} and its experimental evaluation. Provided resources include the source code, dataset/model names, config files, and tools used in the experimental framework. Comprehensive instructions for deployment and carrying out the experiments are available within the linked GitHub repositories~\cite{notebook_os_github_repo, notebook_os_dashboard_github_repo}.

\subsection{Artifact Check-List (Meta-Information)}

{\small
\begin{itemize}
  \item {\bf Model: } {\texttt{\small{VGG-16, ResNet-18, Inception v3, BERT, GPT-2, Deep Speech 2}}}.
  \item {\bf Data set: } CIFAR-10, CIFAR-100, Tiny ImageNet, IMDb Large Movie Reviews, CoLA, LibriSpeech.
  \item {\bf Run-time environment: } AWS, Linux, Ubuntu, WSL 2.
  \item {\bf Hardware: } AWS EC2 virtual machines.
  \item {\bf Metrics: } Interactivity delay, task completion time, number of GPUs provisioned, number of GPUs actively used, profit, profit margin, GPU-hours saved.
  \item {\bf Output: } Numerical statistics.
  \item {\bf Experiments: } Execution of real-world workload trace excerpt on prototype implementation, extension of real-world experiments using simulation study.
  \item {\bf How much disk space is required (approximately)?: } 10's of GB across multiple virtual machines.
  \item {\bf How much time is needed to prepare workflow (approximately)?:} Several hours.
  \item {\bf How much time is needed to complete experiments (approximately)?:} Several hours.
  \item {\bf Publicly available?: } Yes.
  \item {\bf Code licenses (if publicly available)?: } Pending.
  \item {\bf Archived (provide DOI)?: } \href{https://zenodo.org/records/15832099}{\color{linkblue}{10.5281/zenodo.15832098}}.
\end{itemize}
}

\subsection{Description}

\subsubsection{How to Access}

{\proj} can be deployed using a set of Ansible~\cite{ansible} playbooks and shell scripts that we have provided. These tools support deployment across an arbitrary number of AWS EC2 virtual machines, allowing users to customize the scale of their infrastructure. The latest versions of the Ansible playbooks are available in the {\proj} GitHub repository.

\subsubsection{Hardware and Software Dependencies}

{\proj} was developed, tested, and evaluated on WSL 2 (Windows Subsystem for Linux 2) and Ubuntu on AWS EC2. The WSL 2 environment used Ubuntu 22.04.5 LTS (Jammy) with Docker 27.2.0, Go 1.22.9 ({\texttt{\small{linux/amd64}}}), Python 3.12.6, Protoc 27.2, and CUDA 12.7 (via NVIDIA driver v566.36, NVML v565.77, and NVIDIA-SMI v565.77.01). The host OS was Windows 10 v22H2 (OS build 19045.5487), with WSL 2 kernel version 5.15.167.4-1 and WSLg version 1.0.65. %
The AWS EC2 environment ran Ubuntu 24.04.1 LTS (Noble) with Docker 27.3.1 (build ce12230) with matching versions of Golang, Python, and Protoc. Currently, the {\texttt{\small{glibc}}} version in the Global and Local Scheduler Docker containers is {\texttt{\small{ldd (Debian GLIBC 2.36-9+deb12u7) 2.36}}}; within the Jupyter Docker container, it is {\texttt{\small{ldd (Ubuntu GLIBC 2.35-0ubuntu3.8) 2.35}}}.

\noindent\textbf{Remote Storage.} {\proj} persists intermediate workload data to remote storage and currently supports AWS S3 (recommended), Redis and HDFS. For S3, one must create an AWS IAM role with access to the necessary AWS S3 buckets. When provisioning the AWS EC2 virtual machines, be sure that they are assigned an AWS IAM role with access to the AWS S3 buckets intended for use by the distributed kernels.

\subsubsection{Models and Datasets}

To ensure a thorough evaluation of {\proj} and the various other baselines, we integrated support for a variety of different deep learning models and datasets across several different application domains. Table~\ref{tab:datasets_and_models} displays the complete set of models and datasets. 

\subsection{Installation}

{\proj} is deployed on AWS EC2 instances using provided Ansible playbooks (in \texttt{\small{setup/ansible}}~\cite{notebook_os_github_repo}). After provisioning VMs with IAM roles granting S3 access (see the \href{https://docs.aws.amazon.com/AmazonS3/latest/userguide/security_iam_service-with-iam.html}{\color{linkblue}{AWS documentation}}), the playbooks automate dependency installation, patching, and experiment orchestration across the cluster. The source code, as well as additional documentation, is available in the following GitHub repositories:
\begin{enumerate}
    \item \href{https://github.com/ds2-lab/NotebookOS/}{\color{linkblue}{{\proj} source code (primary artifact).}}
    \item \href{https://github.com/ds2-lab/NotebookOS-Dashboard}{\color{linkblue}{Administrative dashboard and workload orchestrator for {\proj}.}}
\end{enumerate}

\subsubsection{Multi-Node Deployments}

The recommended way to deploy {\proj} across multiple AWS EC2 virtual machines is using the provided Ansible playbooks (located in the {\texttt{\small{setup/ansible}}} directory). Before running any playbooks, there are a few configuration-related steps that must be performed. First, create a file called ``{\texttt{\small{all.yaml}}}'' in the {\texttt{\small{setup/ansible/group\_vars}}} directory. The ``{\texttt{\small{all.template.yaml}}}'' file is provided as a starting point. There are five configuration parameters that must be specified explicitly:
\begin{itemize}
    \item \textbf{{\texttt{\small{ansible\_ssh\_private\_key\_file}}}}: Path to a private SSH key on the computer used to run the Ansible playbook(s). Used to enable access to the other VMs in the {\proj} cluster.
    \item \textbf{{\texttt{\small{private\_key\_to\_upload}}}}: Path to a private SSH key on the computer used to run the Ansible playbook(s). This SSH key will be uploaded to the VMs in the {\proj} cluster to enable SSH connectivity between them. This is useful because you may want to run some scripts or Ansible playbooks from one of the VMs once {\proj} has been deployed.
    \item \textbf{{\texttt{\small{public\_key\_to\_upload}}}}: Path to a public SSH key on the computer used to run the Ansible playbook(s). This SSH key will be uploaded to the VMs in the {\proj} cluster to enable SSH connectivity between them. This is useful because you may want to run some scripts or Ansible playbooks from one of the VMs once {\proj} has been deployed.
    \item \textbf{{\texttt{\small{gitbranch}}}}: the branch of the {\proj} GitHub repository to use when deploying {\proj}. You can default to using ``main''.
    \item \textbf{{\texttt{\small{git\_personal\_access\_token}}}}: \href{https://docs.github.com/en/authentication/keeping-your-account-and-data-secure/managing-your-personal-access-tokens}{\color{linkblue}{GitHub personal access token (PAT)}} with read access to the {\proj} GitHub repository (or the fork of the {\proj} source code that is being used).
\end{itemize}

To use Docker images built manually (rather than the provided images), there are several configuration parameters that must be changed. Please consult the documentation in the GitHub repository for additional details.

Once the ``{\texttt{\small{all.yaml}}}'' file has been created in the correct directory (i.e., ``{\texttt{\small{setup/ansible/group\_vars}}}''), you can begin deploying {\proj}. First, run the playbook to create the Docker Swarm cluster:
\begin{lstlisting}[style=bash_cmd_style]
  $ ansible-playbook -i inventory_file.ini create\_docker\_swarm\_cluster.yaml --tags ``swarm''
\end{lstlisting}

Next, deploy the Traefik~\cite{traefik} Docker Stack onto the Docker Swarm cluster. Traefik is an open source reverse proxy and ingress controller that {\proj} uses to route external web traffic to the appropriate internal component. Traefik can be deployed onto the Docker Swarm cluster by executing the following command:
\begin{lstlisting}[style=bash_cmd_style]
  $ ansible-playbook -i inventory_file.ini redeploy\_traefik\_docker\_stack.yaml
\end{lstlisting}

Finally, deploy the {\proj} Docker Stack onto the Docker Swarm cluster:
\begin{lstlisting}[style=bash_cmd_style]
  $ ansible-playbook -i inventory_file.ini deploy_distributed_notebook\_docker\_stack.yaml
\end{lstlisting}

If desired, verbose Ansible logging can be enabled by setting the {\texttt{\small{ANSIBLE\_STDOUT\_CALLBACK}}} environment variable to ``{\texttt{\small{debug}}}''. For example:
\begin{lstlisting}[style=bash_cmd_style]
  $ ANSIBLE_STDOUT_CALLBACK=debug ansible-playbook -i inventory_file.ini deploy\_distributed\_notebook\_docker\_stack.yaml
\end{lstlisting}

\subsubsection{Single-Node and Development Deployments}

For development, {\proj} supports Docker Compose~\cite{docker_compose}. Run \texttt{\small setup/install.sh} to install host dependencies. A template for the Docker Compose \texttt{\small yml} file is provided in the \texttt{\small deploy/docker-WSL2/} directory; generate it via the ``generate-docker-compose-file.sh'' script. Helper scripts and usage details are documented in the directory’s README. Once the {\texttt{\small docker-compose.yml}} file is generated, you can deploy {\proj} via Docker Compose using the following command:

\begin{lstlisting}[style=bash_cmd_style]
  $ docker compose up -d --build --scale daemon=4
\end{lstlisting}

In order for {\proj} to operate correctly with 3 replicas per distributed kernel, the minimum value of the ``{\texttt{\small{--scale daemon=}}}'' argument is 3; however, the recommended minimum is 4 to enable kernel replica migrations to occur.

When deploying {\proj} for development, we recommend specifying a few additional configuration parameters. First, the system expects that the core dumps will be written to a {\texttt{\small{/cores}}} directory. We recommend mounting a Docker volume so that data written to the {\texttt{\small{/cores}}} directory are persisted and can be accessed later. We also recommend deploying Dozzle~\cite{dozzle} for easier log monitoring.

\begin{lstlisting}[style=bash_cmd_style]
  $ docker run --name dozzle -d \ 
    -v /var/run/docker.sock:/var/run/docker.sock:ro \ 
    -p 7744:8080 amir20/dozzle:latest
\end{lstlisting}

\subsubsection{Running the Dashboard Locally or Independently}

After cloning the dashboard repository, execute the following command from within the {\texttt{\small{driver-frontend}}} directory: 
\begin{lstlisting}[style=bash_cmd_style]
  $ npm install && npm run start:dev
\end{lstlisting}

Note that unless the above command is modified, the front-end server will become the active process for the terminal.

Next, to run the backend server, execute the following command from within the {\texttt{\small{driver-backend}}} directory: 
\begin{lstlisting}[style=bash_cmd_style]
  $ make run-server
\end{lstlisting}

Note that unless the avove command is modified, the back-end server will become the active process for the terminal.

\subsubsection{AWS Configuration}

To deploy {\proj} on AWS, designate one internet-accessible EC2 VM with a public IPv4 address as the ``primary VM''. This VM, which should be named {\texttt{\small{``Jupyter NaaS Leader''}}}, hosts {\proj}’s Traefik service, which routes external traffic (e.g., to the workload dashboard). Use the {\texttt{\small{generate\_inventory\_file\_aws.py}}} file in the {\texttt{\small{setup/ansible/}}} directory to generate an {\texttt{\small{inventory.ini}}} from {\texttt{\small{inventory\_file.template.ini}}}. The script requires the primary VM to be named as above, and all other VMs in the {\proj} cluster should be tagged with key {\texttt{\small{``swarm''}}} and value {\texttt{\small{``follower''}}}. It assumes Ubuntu VMs with the {\texttt{\small{``ubuntu''}}} root username.

\subsection{Experiment Workflow}

The experiments for {\proj} and the other baselines are primarily orchestrated using {\proj}'s administrative dashboard~\cite{notebook_os_dashboard_github_repo}. The dashboard provides a browser-based interface to view and manage {\proj} itself, as well as workloads running on {\proj}.

\subsection{Evaluation and Expected Results}

The results presented in Section~\ref{sec:evaluation} correspond to the execution of a real-world workload trace on {\proj}. We are working to open-source our workload trace. Executing the same workload on {\proj} multiple times will generate approximately the same results, with small differences resulting from scheduling decisions and other random factors.

\section{Limited GPU Resource Availability}

The demand for GPUs has surged in recent years, driven by the rapid growth of machine learning, deep learning, and high-performance computing applications~\cite{gpu_reportsinsights, scaling_market_structure_ai}. However, cloud providers such as Amazon Web Services (AWS), Google Cloud, and Microsoft Azure are struggling to meet this increasing demand due to limited GPU resources~\cite{skypilot_nsdi23, gpu_shortage_euromlsys24, data_center_risk, scaling_market_structure_ai}. This scarcity is further compounded by global supply chain challenges and the growing competition from sectors like gaming, cryptocurrency mining, and scientific research, all of which intensify the pressure on cloud infrastructure providers to allocate GPU resources efficiently~\cite{read_these_2021, sweney_global_2021, scaling_market_structure_ai}. The low availability of GPUs serves as further motivation of {\proj}.

\section{Implementation}
\label{sec:appendix-implementation}

To demonstrate {\proj}'s efficacy, we implemented a fully functional prototype Jupyter Notebook platform. 
{\proj} reuses the existing IPython messaging protocol used by standard Jupyter clients and kernels, ensuring compatibility with all Jupyter clients. 
{\proj} uses the Jupyter Server from the official \texttt{\small{base-notebook}} Docker image. {\proj} uses a custom kernel provisioner, referred to as the \texttt{\small{GatewayProvisioner}}, as well as a custom \texttt{\small{SessionManager}} and \texttt{\small{KernelManager}}. All custom components were implemented using the official Jupyter Server API extension methods, building on the default implementations provided by Jupyter Server.  
{\proj} also implements a custom Jupyter Kernel Provisioner, \texttt{\small{GatewayProvisioner}} to integrate directly with standard/vanilla Jupyter deployments. Jupyter Kernel Provisioners enable third parties to manage the life-cycle of a kernel's runtime environment. 
{\proj}'s Global Scheduler and Local Schedulers communicate using gRPC, while kernel-replica-related messages use ZMQ.

\begin{table}[ht]
\caption{Lines of code required for the development and evaluation of {\proj} by its components.} 
\vspace{-8pt}
\label{tab:loc-table}
\scalebox{0.76}{ %
    \begin{tabular}{@{}lr|lr@{}}
      
      \hhline{|----|}
      
      \rowcolor[HTML]{d6eaf8} 
      {\bf Component} & {\bf LoC} & {\bf Component} & {\bf LoC} \\
      
      \hhline{|----|}
      
      {Simulator (Go/Python)} & 40,395 & {Prototype (Go/Python)} & 175,239 \\
      
      \rowcolor[HTML]{EFEFEF} 
      Cluster Dashboard (TypeScript) & 23,038 & Misc. scripts & 4,322 \\
      
      Workload Orchestrator (Go) & 43,043 & \cellcolor[HTML]{9AFF99}{\textbf{Total:}} & \cellcolor[HTML]{9AFF99}285,037 
    \end{tabular}
}
\vspace{-8pt}
\end{table}

We have implemented {\proj} 
in approximately 282k lines of code  over roughly two person-years. 
See Table~\ref{tab:loc-table} for a summary of our implementation efforts.

\vspace{-3pt}
\section{Simulation Study: Resource Usage}
\label{subsec:eval_sim_study_resource_usage}

\begin{figure}[ht]
\begin{center}
\subfigure[Cluster-wide allocatable GPUs.] {
\includegraphics[height=0.22\textwidth]{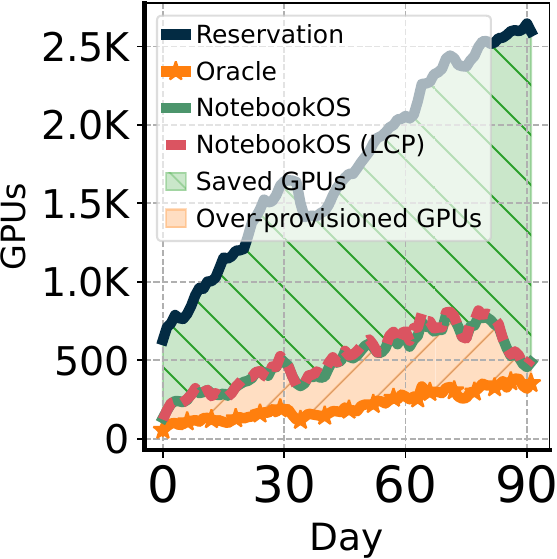}
\label{fig:sim_study_gpu_usage_shaded}
}
\hspace{-10pt}
\subfigure[GPU usage ratio.] {
\includegraphics[height=0.22\textwidth]{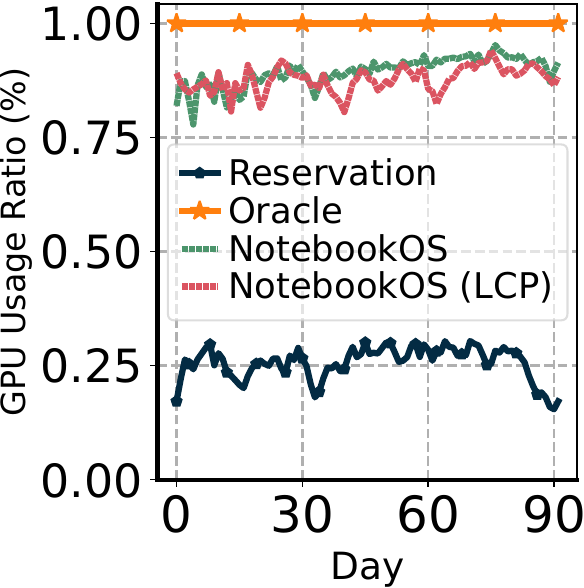}
\label{fig:sim_study_gpu_usage_ratio}
}
\vspace{-12pt}
\Description[Simulated GPU usage timeline over a 90-day snippet of the {\platformtrace} spanning June through August.]{Simulated GPU usage timeline over a 90-day snippet of the {\platformtrace} spanning June through August.}
\caption{Simulated GPU usage timeline over a 90-day snippet of the {\platformtrace} spanning June through August.} 
\label{fig:sim_study_gpu_usage_shaded_and_ratio}
\vspace{-5pt}
\end{center}
\end{figure}

Our simulation study of the 3-month trace confirms the significantly improved GPU efficiency over {\reservation} (Figure~\ref{fig:sim_study_gpu_usage_ratio}). This is because {\proj} oversubscribes server resources and consequently requires fewer servers to be provisioned. 
Figure~\ref{fig:sim_study_gpu_usage_ratio} plots the ratio of allocatable GPUs that are actively utilized by kernel containers during the simulated 3-month workload. 
{\proj} uses a significantly higher fraction of the available GPUs compared to the \texttt{\small{reservation}} baseline, again illustrating {\proj}'s resource efficiency.

\section{Detailed Latency \& Overhead Breakdown}

\begin{figure}[t]
\centering
\includegraphics[width=0.485\textwidth]{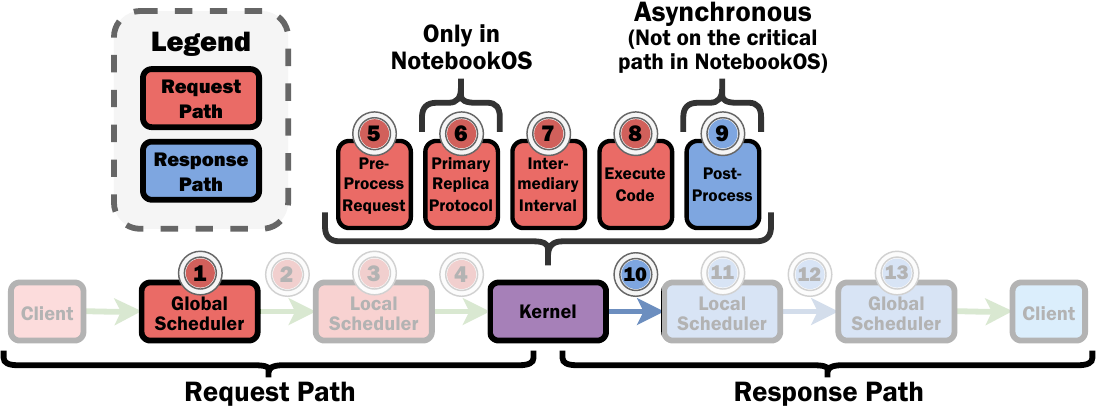}
\vspace{-18pt} 
\Description[Workflow steps of notebook cell execution requests in {\proj} and other baselines.]{Workflow steps of notebook cell execution requests in {\proj} and other baselines.}
\caption{Workflow steps of notebook cell execution requests in {\proj} and other baselines.} 
\label{fig:execution_steps}
\vspace{-8pt} 
\end{figure}

\begin{figure}[t]
\centering
\includegraphics[width=0.475\textwidth]{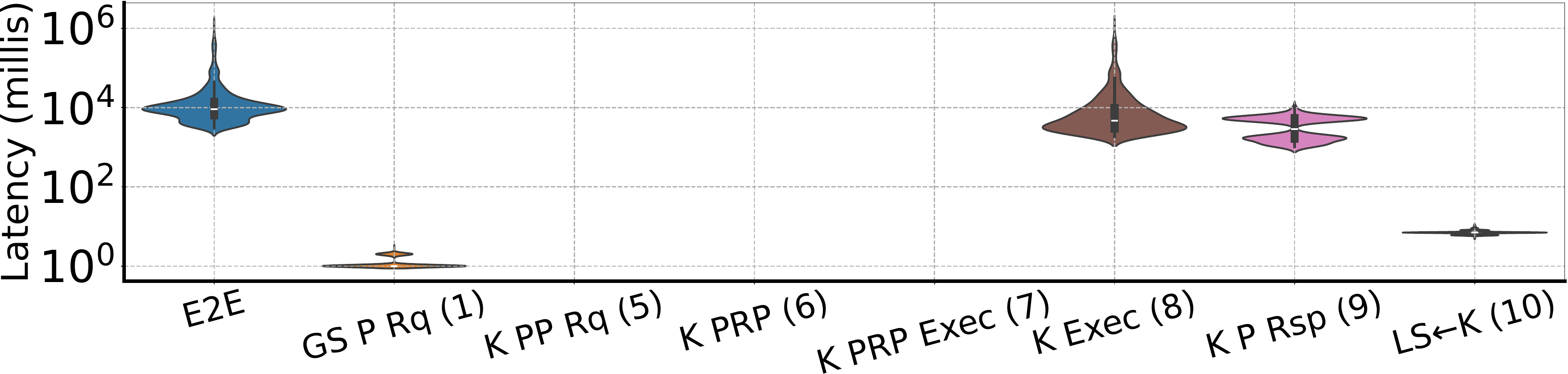}
\vspace{-18pt} 
\Description[Detailed end-to-end latency breakdown of execute requests messages processed by {\reservation}.]{Detailed end-to-end latency breakdown of execute requests messages processed by {\reservation}.}
\caption{Detailed end-to-end latency breakdown of execute requests messages processed by {\reservation}.} 
\label{fig:detailed_latency_reservation}
\vspace{-8pt} 
\end{figure}

\begin{figure}[t]
\centering
\includegraphics[width=0.475\textwidth]{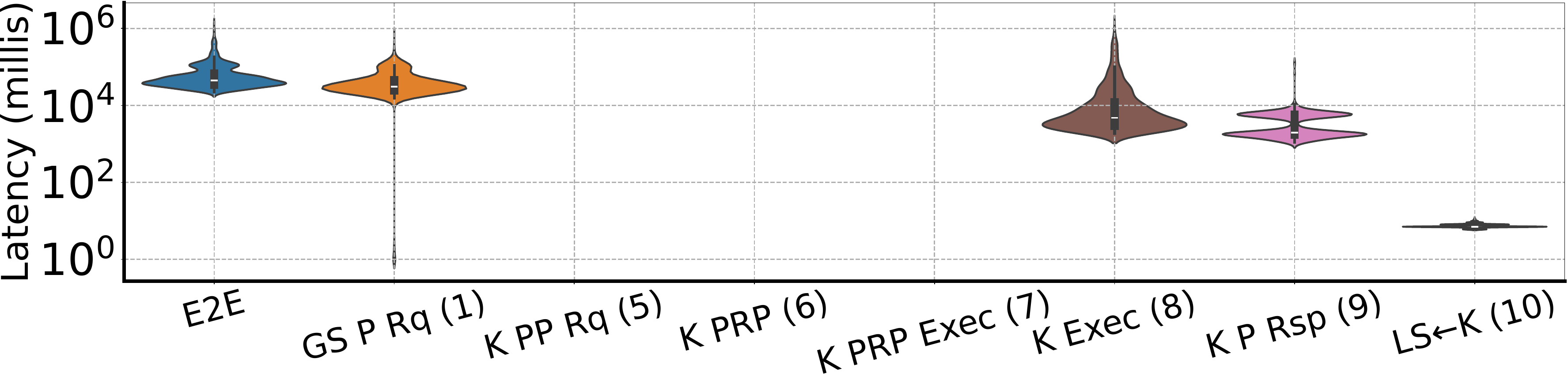}
\vspace{-18pt} 
\Description[Detailed end-to-end latency breakdown of execute requests messages processed by {\FCFS}.]{Detailed end-to-end latency breakdown of execute requests messages processed by {\FCFS}.}
\caption{Detailed end-to-end latency breakdown of execute requests messages processed by {\FCFS}.} 
\label{fig:detailed_latency_fcfs}
\vspace{-8pt} 
\end{figure}

\begin{figure}[t]
\centering
\includegraphics[width=0.475\textwidth]{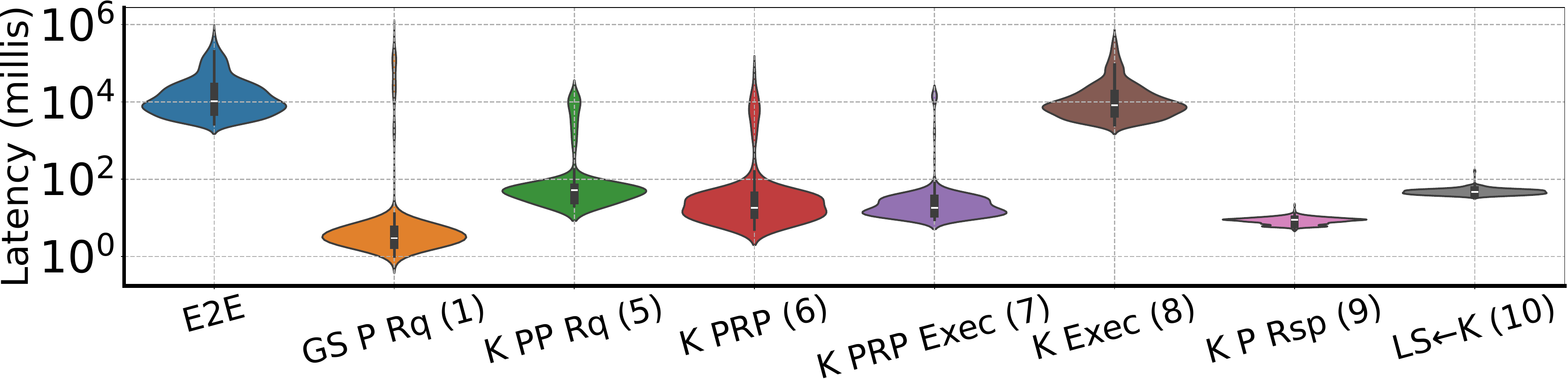}
\vspace{-18pt} 
\Description[Detailed end-to-end latency breakdown of execute requests messages processed by {\proj}.]{Detailed end-to-end latency breakdown of execute requests messages processed by {\proj}.}
\caption{Detailed end-to-end latency breakdown of execute requests messages processed by {\proj}.} 
\label{fig:detailed_latency_europa}
\vspace{-8pt} 
\end{figure}

\begin{figure}[t]
\centering
\includegraphics[width=0.475\textwidth]{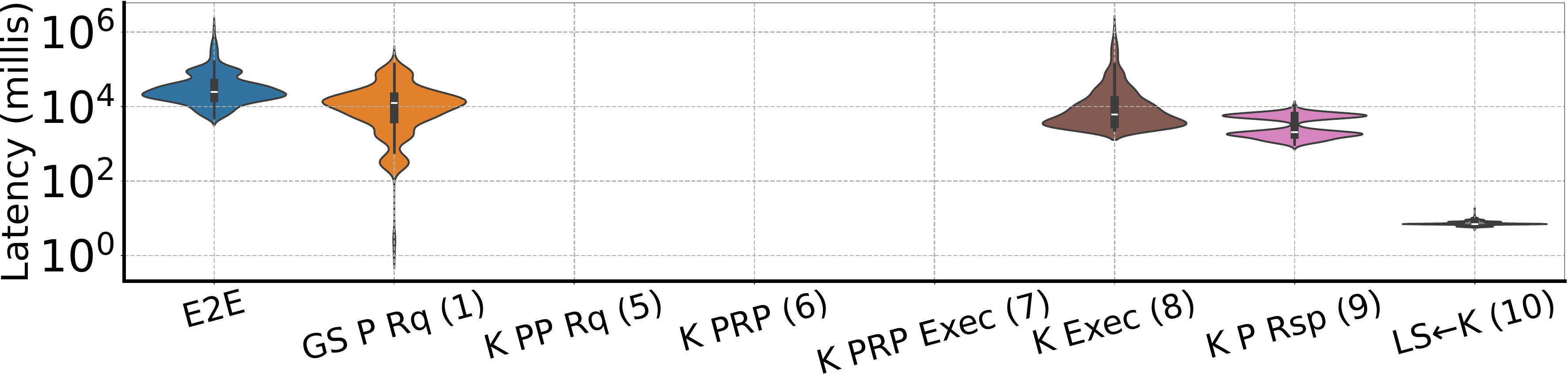}
\vspace{-18pt} 
\Description[Detailed end-to-end latency breakdown of execute requests messages processed by {\projlcp}.]{Detailed end-to-end latency breakdown of execute requests messages processed by {\projlcp}.}
\caption{Detailed end-to-end latency breakdown of execute requests messages processed by {\projlcp}.} 
\label{fig:detailed_latency_europa_lcp}
\vspace{-8pt} 
\end{figure}

Figure~\ref{fig:execution_steps} presents an overview of the individual steps along the critical path of execution requests within {\proj} and other baselines. 
Some steps are unique to {\proj} and do not exist in other baselines, and therefore, they have zero delay in other baselines. 

Step~\circled{1} corresponds to the processing performed by the Global Scheduler when an execution request is received. This may include on-demand docker container provisioning, queuing delays, and the overhead associated with making placement decisions. Step~\circled{2} is the latency of the network hop between the Global Scheduler and the Local Scheduler. Step~\circled{3} is any processing performed by the Local Scheduler. Step~\circled{4} is the latency of the network hop between the Local Scheduler and the kernel replica. Step~\circled{5} includes pre-processing performed by the kernel replica, which may involve extracting metadata. Step~\circled{6}, which only occurs while using {\proj}, is the executor replica selection protocol. Step~\circled{7} includes the time between the selection of the executor replica and the beginning of code execution, and Step~\circled{8} is the execution of the user-submitted cell task by the kernel replica. Step~\circled{9} involves post-processing by the kernel replica, which in {\proj} may include state synchronization via Raft SMR and writing large objects to the Distributed Data Store. 
In {\proj}, this step is \emph{asynchronous} and does not affect the user experience. 
The remaining steps largely just involve forwarding the response back to the client. Steps that appear more lightly-colored in Figure~\ref{fig:execution_steps} are omitted from the detailed request latency figures discussed next. These steps are omitted because their latency is near zero for all baselines.

Figure~\ref{fig:detailed_latency_reservation} presents a detailed breakdown and overview of the steps along the critical path of execution requests as observed by the {\reservation} baseline. Execution requests processed by the {\reservation} baseline spend the most time in Step~\circled{8}, which corresponds to the kernel replica executing the user-submitted code. Step 9 also incurs some latency, as the kernel replica persists the updated state to remote storage during this step.

Figure~\ref{fig:detailed_latency_fcfs} presents a detailed breakdown and overview of the steps along the critical path of execution requests as observed by the {\FCFS} baseline. Execution requests processed by the {\FCFS} baseline encounter significant delays during Step~\circled{1}, which includes queuing delays and on-demand docker container provisioning. The same is true for {\projlcp}, though {\projlcp} observes shorter latencies during this step because many requests can simply be processed by an existing, pre-warmed container.

Figure~\ref{fig:detailed_latency_europa} presents a detailed breakdown and overview of the steps along the critical path of execution requests as observed by {\proj}. {\proj} incurs slightly increased overhead in many steps compared to the other baselines; however, the small amount of additional overhead incurred in these steps does not outweigh the savings (in terms of interactivity and task completion time) enabled by {\proj}. {\proj} primarily incurs additional overhead when execution requests are being processed by the kernel replicas. This is because the executor replica selection protocol must be performed (Step~\circled{6} in Figure~\ref{fig:execution_steps}). This protocol typically takes tens of milliseconds at most, however, and so it does not contribute significantly to the overall end-to-end latency.  

\subsection{Active Sessions \& Training Events}

\begin{figure}[t]
\centering
\includegraphics[width=0.475\textwidth]{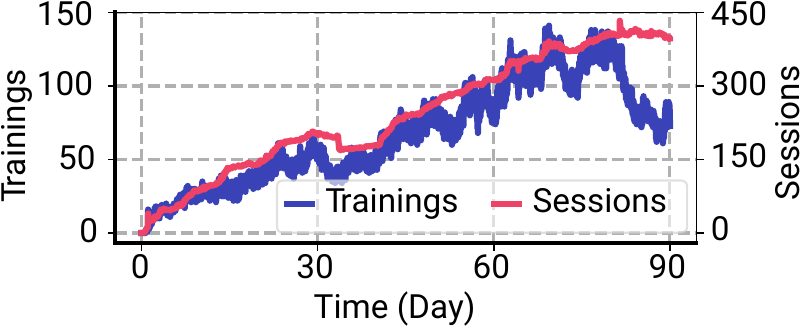}
\vspace{-18pt} 
\Description[The number of active user-submitted trainings and active user sessions during the full summer portion (i.e., June, July, and August) of the {\idltplatform} workload trace.]{The number of active user-submitted trainings and active user sessions during the full summer portion (i.e., June, July, and August) of the {\idltplatform} workload trace.}
\caption{The number of active user-submitted trainings and active user sessions during the full summer portion (i.e., June, July, and August) of the {\idltplatform} workload trace.} 
\label{fig:active_sessions_and_trainings_full_summer}
\vspace{-14pt} 
\end{figure}

Figure~\ref{fig:active_sessions_and_trainings_full_summer} presents a timeline plot showing the number of active sessions during the ``summer portion'' of the {\idltplatform} workload trace. Specifically, the number of active sessions is plotted on the secondary (i.e., right) y-axis. The summer portion includes data from the entirety of the months of June, July, and August. Initially, there are 0 active user sessions. By the end of June, July, and August, there are 206, 312, and 397 active user sessions, respectively. For June, the mean and median number of active user sessions are 115 and 130, respectively. For July, the mean and median number of active user sessions are 233 and 234, respectively. Finally, for August, the mean and median number of active user sessions are 379 and 385, respectively. The maximum number of active sessions at any given time throughout the summer {\idltplatform} workload trace is 433.

Figure~\ref{fig:active_sessions_and_trainings_full_summer} also presents a timeline plot showing the number of user-submitted training events being processed during the ``summer portion'' of the {\idltplatform} workload trace. Specifically, the number of active sessions is plotted on the primary (i.e., left) y-axis. By the end of June, July, and August, there are 58, 83, and 73 active user-submitted trainings, respectively. The mean and median number of active user-submitted trainings for the entire summer are 67.63
and 68, respectively. For June, the mean and median number of active user-submitted trainings are 31 and 30, respectively. For July, the mean and median number of active user-submitted trainings are 65 and 67, respectively. Finally, for August, the mean and median number of active user-submitted trainings are 105 and 108, respectively. The maximum number of active user-submitted trainings at any given time throughout the summer {\platformtrace} is 141.

\end{document}